%% file: ms.tex
\documentclass[twocolumn]{aastex61}

\usepackage{epstopdf}
\usepackage{amsmath, amsthm, amssymb}
\usepackage{fancybox}
\usepackage{graphicx}
\usepackage{color}
\usepackage{epsfig}

\usepackage{epsfig,natbib}
\usepackage{graphicx}
\usepackage{amsmath}

\def\apjl{ApJL}
\def\apjs{ApJS}
\def\ao{Appl.~Opt.}
\def\aap{A\&A}
 
\DeclareMathAlphabet{\mathsc}{OT1}{cmr}{m}{sc}
\def\testbx{bx}%
\DeclareRobustCommand{\ion}[2]{%
\relax\ifmmode
\ifx\testbx\f@series
{\mathbf{#1\,\mathsc{#2}}}\else
{\mathrm{#1\,\mathsc{#2}}}\fi
\else\textup{#1\,{\mdseries\textsc{#2}}}%
\fi}

\newcommand{\Hii} {\ion{H}{ii}}
\newcommand{\ha} {\mbox{H$\alpha$}}

\newcommand{\Hei} {\ion{He}{i}}

\newcommand{\Nii} {\ion{N}{ii}}

\newcommand{\Oi} {\ion{O}{i}}
\newcommand{\Oii} {\ion{O}{ii}}

\newcommand{\Nai} {\ion{Na}{i}}

\newcommand{\Siii} {\ion{Si}{ii}}

\newcommand{\Caii} {\ion{Ca}{ii}}
\newcommand{\Scii} {\ion{Sc}{ii}}

\newcommand{\Feii} {\ion{Fe}{ii}}

\newcommand{\snfull}{Gaia17biu/SN~2017egm}
\newcommand{\sn}{Gaia17biu}
\newcommand{\host}{NGC~3191}
\newcommand{\EpEpoch}{JD~2,457,926.3}

\newcommand{\synow}{\textsc{synow}}

\newcommand{\msun}{\mbox{M$_{\odot}$}}

\newcommand{\kms}{\mbox{$\rm{\,km\,s^{-1}}$}}

\newcommand{\ld}{\mbox{$\lambda$}}

\newcommand{\fdegr}{\ensuremath{.\!\!^{\circ}}} 
\newcommand{\lunits}{\,\ensuremath{\mathrm{erg\,s^{-1}\,Hz^{-1}}}}  
\newcommand{\mujyb}{\,\mbox{\ensuremath{\mathrm{\mu Jy/beam}}}}      
\newcommand{\hh}{\ensuremath{\mathrm{^h}}}                           
\newcommand{\mm}{\ensuremath{\mathrm{^m}}}                           

\begin{document} 
 
\title{\snfull\ in NGC 3191: The closest hydrogen-poor superluminous supernova to date is in a ``normal'', massive, metal-rich spiral galaxy} 
\correspondingauthor{Subo Dong, Subhash Bose}
\email{dongsubo@pku.edu.cn, email@subhashbose.com}
\author{Subhash Bose}\affil{Kavli Institute for Astronomy and Astrophysics, Peking University, Yi He Yuan Road 5, Hai Dian District, Beijing 100871, China}
\author{Subo Dong}\affil{Kavli Institute for Astronomy and Astrophysics, Peking University, Yi He Yuan Road 5, Hai Dian District, Beijing 100871, China}
\author{A. Pastorello}\affil{INAF-Osservatorio Astronomico di Padova, Vicolo dell'Osservatorio 5, I-35122 Padova, Italy}
	\author{Alexei V. Filippenko} \affil{Department of Astronomy, University of California, Berkeley, CA 94720-3411, USA} \affil{Miller Senior Fellow, Miller Institute for Basic Research in Science, University of California, Berkeley, CA 94720, USA.}																						
	\author{C. S. Kochanek} \affil{Department of Astronomy, The Ohio State University, 140 W. 18th Avenue, Columbus, OH 43210, USA.} \affil{Center for Cosmology and AstroParticle Physics (CCAPP), The Ohio State University, 191 W. Woodruff Avenue, Columbus, OH 43210, USA.}								
	\author{Jon Mauerhan} \affil{Department of Astronomy, University of California, Berkeley, CA 94720-3411, USA}																								
	\author{C. Romero-Ca\~nizales} \affil{Millennium Institute of Astrophysics, Santiago, Chile.} \affil{N\'ucleo de Astronom\'ia de la Facultad de Ingenier\'ia y Ciencias, Universidad Diego Portales, Av. Ej \'ercito 441, Santiago, Chile}																								
	\author{Thomas G. Brink} \affil{Department of Astronomy, University of California, Berkeley, CA 94720-3411, USA}																								
	\author{Ping Chen} \affil{Kavli Institute for Astronomy and Astrophysics, Peking University, Yi He Yuan Road 5, Hai Dian District, Beijing 100871, China}																								
	\author{J. L. Prieto} \affil{N\'ucleo de Astronom\'ia de la Facultad de Ingenier\'ia y Ciencias, Universidad Diego Portales, Av. Ej \'ercito 441, Santiago, Chile}	 \affil{Millennium Institute of Astrophysics, Santiago, Chile.}																				
	\author{R. Post} \affil{Post Observatory, Lexington, MA 02421}																								
	\author{Christopher Ashall} \affil{Department of Physics, Florida State University, 77 Chieftain Way, Tallahassee, FL 32306, USA}																								
	\author{Dirk Grupe} \affil{Department of Earth and Space Science, Morehead State University, 235 Martindale Dr., Morehead, KY 40351, USA}																								
	\author{L. Tomasella} \affil{INAF-Osservatorio Astronomico di Padova, Vicolo dell'Osservatorio 5, I-35122 Padova, Italy}																								
	\author{Stefano Benetti} \affil{INAF-Osservatorio Astronomico di Padova, Vicolo dell'Osservatorio 5, I-35122 Padova, Italy}																								
	\author{B. J. Shappee} \affil{Carnegie Observatories, 813 Santa Barbara Street, Pasadena, CA 91101, USA} \affil{Hubble Fellow} \affil{Carnegie-Princeton Fellow}																								
	\author{K. Z. Stanek} \affil{Department of Astronomy, The Ohio State University, 140 W. 18th Avenue, Columbus, OH 43210, USA.} \affil{Center for Cosmology and AstroParticle Physics (CCAPP), The Ohio State University, 191 W. Woodruff Avenue, Columbus, OH 43210, USA.}																								
	\author{Zheng Cai} \affil{UCO/Lick Observatory, University of California at Santa Cruz, Santa Cruz, CA, 95064}																								
	\author{E. Falco} \affil{Harvard-Smithsonian Center for Astrophysics, 60 Garden Street, Cambridge, MA 02138, USA.}																								
	\author{Peter Lundqvist} \affil{Department of Astronomy and The Oskar Klein Centre, AlbaNova University Center, Stockholm University, SE-10691 Stockholm, Sweden}																								
	\author{Seppo Mattila} \affil{Tuorla Observatory, Department of Physics and Astronomy, University of Turku, Väisäläntie 20, FI-21500 Piikkiö, Finland}																								
	\author{Robert Mutel} \affil{Department of Physics and Astronomy, University of Iowa}																			
	\author{Paolo Ochner} \affil{INAF-Osservatorio Astronomico di Padova, Vicolo dell'Osservatorio 5, I-35122 Padova, Italy} \affil{Dipartimento di Fisica e Astronomia, Universit`a di Padova, via Marzolo 8, I-35131 Padova, Italy}						
	\author{David Pooley} \affil{Trinity University, Department of Physics \& Astronomy, One Trinity Place, San Antonio, TX 78212}																								
	\author{M. D. Stritzinger} \affil{Department of Physics and Astronomy, Aarhus University, Ny Munkegade 120, DK-8000 Aarhus C, Denmark}																								
	\author{S. Villanueva Jr.} \affil{Department of Astronomy, The Ohio State University, 140 W. 18th Avenue, Columbus, OH 43210, USA.}																								
	\author{WeiKang Zheng} \affil{Department of Astronomy, University of California, Berkeley, CA 94720-3411, USA}																								
	\author{R. J. Beswick} \affil{Jodrell Bank Centre for Astrophysics \& e-MERLIN, School of Physics and Astronomy, The University of Manchester, Manchester, M13 9PL, UK}																								
	\author{Peter J. Brown} \affil{George P. and Cynthia Woods Mitchell Institute for Fundamental Physics \& Astronomy, Texas A. \& M. University, Department of Physics and Astronomy, 4242 TAMU, College Station, TX 77843, USA}																								
	\author{E. Cappellaro} \affil{INAF-Osservatorio Astronomico di Padova, Vicolo dell'Osservatorio 5, I-35122 Padova, Italy}																								
	\author{Scott Davis} \affil{Department of Physics, Florida State University, 77 Chieftain Way, Tallahassee, FL 32306, USA}																								
	\author{Morgan Fraser} \affil{School of Physics, O'Brien Centre for Science North, University College Dublin, Belfield, Dublin 4}
	\author{Thomas de Jaeger} \affil{Department of Astronomy, University of California, Berkeley, CA 94720-3411, USA}																								
	\author{N. Elias-Rosa} \affil{INAF-Osservatorio Astronomico di Padova, Vicolo dell'Osservatorio 5, I-35122 Padova, Italy}																								
	\author{C. Gall} \affil{Dark Cosmology Centre, Niels Bohr Institute, University of Copenhagen, Juliane Maries Vej 30, 2100 Copenhagen}																								
	\author{B. Scott Gaudi} \affil{Department of Astronomy, The Ohio State University, 140 W. 18th Avenue, Columbus, OH 43210, USA.}																								
	\author{Gregory J. Herczeg} \affil{Kavli Institute for Astronomy and Astrophysics, Peking University, Yi He Yuan Road 5, Hai Dian District, Beijing 100871, China}																								
	\author{Julia Hestenes} \affil{Department of Astronomy, University of California, Berkeley, CA 94720-3411, USA}																								
	\author{T. W.-S. Holoien} \affil{Department of Astronomy, The Ohio State University, 140 W. 18th Avenue, Columbus, OH 43210, USA.} \affil{Center for Cosmology and AstroParticle Physics (CCAPP), The Ohio State University, 191 W. Woodruff Avenue, Columbus, OH 43210, USA.} \affil{US Department of Energy Computational Science Graduate Fellow}																								
	\author{Griffin Hosseinzadeh} \affil{Las Cumbres Observatory, 6740 Cortona Dr Ste 102, Goleta, CA 93117-5575, USA} \affil{Department of Physics, University of California, Santa Barbara, CA 93106-9530, USA}																								
	\author{E. Y. Hsiao} \affil{Department of Physics, Florida State University, 77 Chieftain Way, Tallahassee, FL 32306, USA}																								
	\author{Shaoming Hu} \affil{Shandong Provincial Key Laboratory of Optical Astronomy and Solar-Terrestrial Environment, Institute of Space Sciences, Shandong University, Weihai 264209, China}																								
	\author{Shin Jaejin} \affil{Astronomy Program, Department of Physics and Astronomy, Seoul National University, Seoul, 151-742, Korea}																								
	\author{Ben Jeffers} \affil{Department of Astronomy, University of California, Berkeley, CA 94720-3411, USA}																								
	\author{R. A. Koff} \affil{Antelope Hills Observatory 980 Antelope DR W Bennett, CO 80102 USA}																								
	\author{Sahana Kumar} \affil{Department of Physics, Florida State University, 77 Chieftain Way, Tallahassee, FL 32306, USA}																								
	\author{Alexander Kurtenkov} \affil{Institute of Astronomy and NAO, Bulgarian Academy of Sciences, 72 Tsarigradsko Shose Blvd., 1784 Sofia, Bulgaria}																								
	\author{Marie Wingyee Lau} \affil{UCO/Lick Observatory, University of California at Santa Cruz, Santa Cruz, CA, 95064}																																									
	\author{Simon Prentice} \affil{Astrophysics Research Institute, Liverpool John Moores University, Liverpool, L3 5RF, UK}																								
	\author{T. Reynolds} \affil{Tuorla Observatory, Department of Physics and Astronomy, University of Turku, Väisäläntie 20, FI-21500 Piikkiö, Finland}
	\author{Richard J. Rudy} \affil{Space Science Applications Laboratory, The Aerospace Corporation}																								
	\author{Melissa Shahbandeh} \affil{Department of Physics, Florida State University, 77 Chieftain Way, Tallahassee, FL 32306, USA}																								
	\author{Auni Somero} \affil{Tuorla Observatory, Department of Physics and Astronomy, University of Turku, Väisäläntie 20, FI-21500 Piikkiö, Finland}																								
	\author{Keivan G. Stassun} \affil{Department of Physics \& Astronomy, Vanderbilt University, 6301 Stevenson Center Ln., Nashville, TN 37235 USA}																								
	\author{Todd A. Thompson} \affil{Department of Astronomy, The Ohio State University, 140 W. 18th Avenue, Columbus, OH 43210, USA.} \affil{Center for Cosmology and AstroParticle Physics (CCAPP), The Ohio State University, 191 W. Woodruff Avenue, Columbus, OH 43210, USA.}																								
	\author{Stefano Valenti} \affil{Department of Physics, University of California, Davis, CA 95616, USA}																								
	\author{Jong-Hak Woo} \affil{Astronomy Program, Department of Physics and Astronomy, Seoul National University, Seoul, 151-742, Korea}																								
	\author{Sameen Yunus} \affil{Department of Astronomy, University of California, Berkeley, CA 94720-3411, USA}																					
\begin{abstract}
Hydrogen-poor superluminous supernovae (SLSNe-I) have been predominantly found in low-metallicity, star-forming dwarf galaxies. Here we identify \snfull\ as an SLSN-I occurring in a ``normal'' spiral galaxy (NGC~3191) in terms of stellar mass (several times $10^{10}$\,M$_\odot$) and metallicity (roughly Solar). At redshift $z=0.031$, \sn\ is also the lowest redshift SLSN-I to date, and the absence of a larger population of SLSNe-I in dwarf galaxies of similar
redshift suggests that metallicity is likely less important to the production of SLSNe-I than previously believed. With the smallest distance and highest apparent brightness for an SLSN-I, we are able to study \sn\ in unprecedented detail. Its pre-peak near-ultraviolet to optical color is similar to that of Gaia16apd and among the bluest observed for an SLSN-I while its peak luminosity ($M_g = -21$\,mag) is substantially lower than Gaia16apd. Thanks to the high signal-to-noise ratios of our spectra, we identify several new spectroscopic features that may help to probe the properties of these enigmatic explosions. We detect polarization at the $\sim 0.5\%$ level that is not strongly dependent on wavelength, suggesting a modest, global departure from spherical symmetry. In addition, we put the tightest upper limit yet on the radio luminosity of an SLSN-I with $<5.4\times10^{26}\,{\rm erg\,s^{-1}\,Hz^{-1}}$ at $10\,{\rm GHz}$, which is almost a factor of 40 better than previous upper limits and one of the few measured at an early stage in the evolution of an SLSN-I. This limit largely rules out an association of this SLSNe-I with known populations of gamma-ray burst (GRB) like central engines.
\end{abstract}
 

\section{Introduction} \label{sec:intro}

\begin{figure*}
	\centering
	\includegraphics[width=0.35\linewidth]{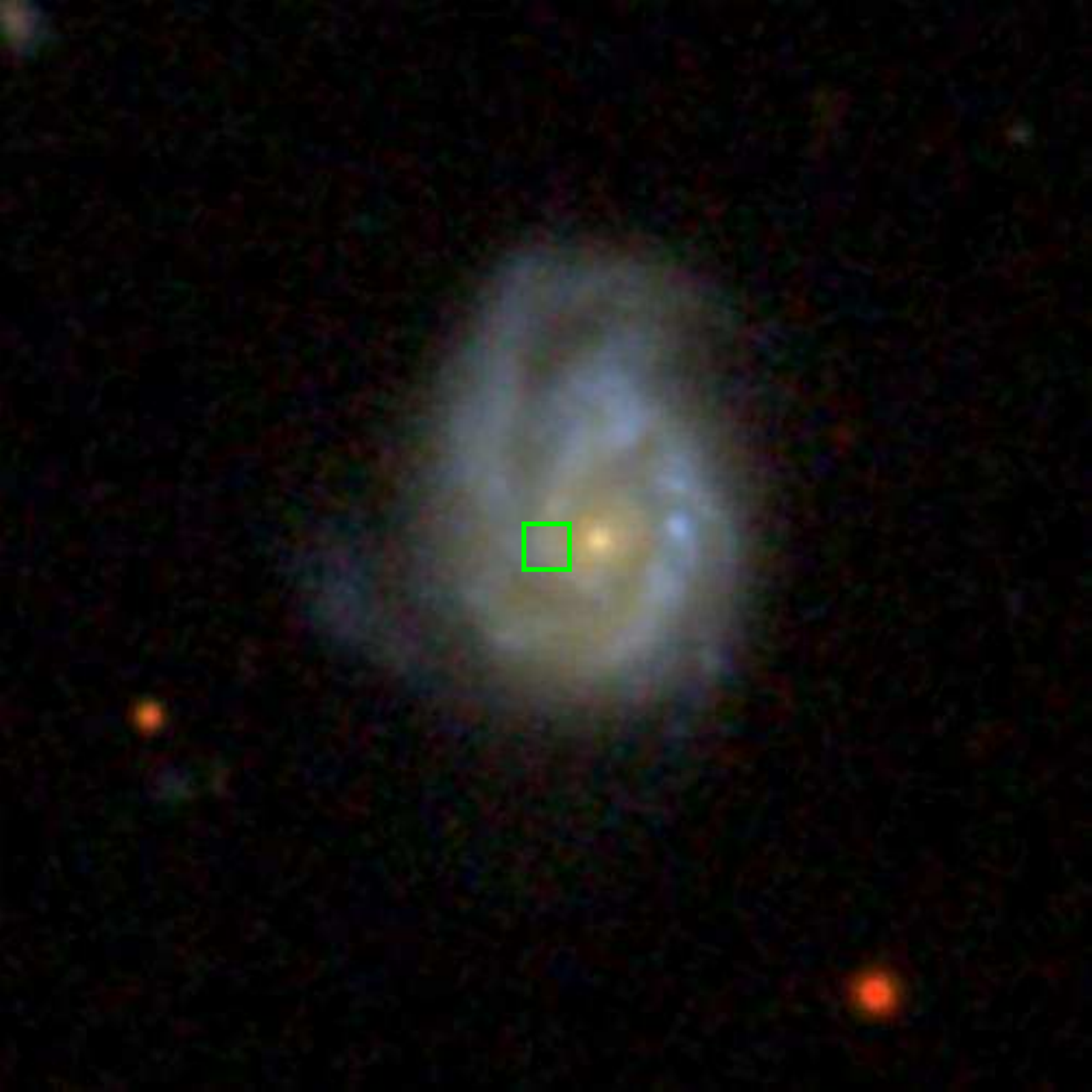}%
	\includegraphics[width=0.35\linewidth]{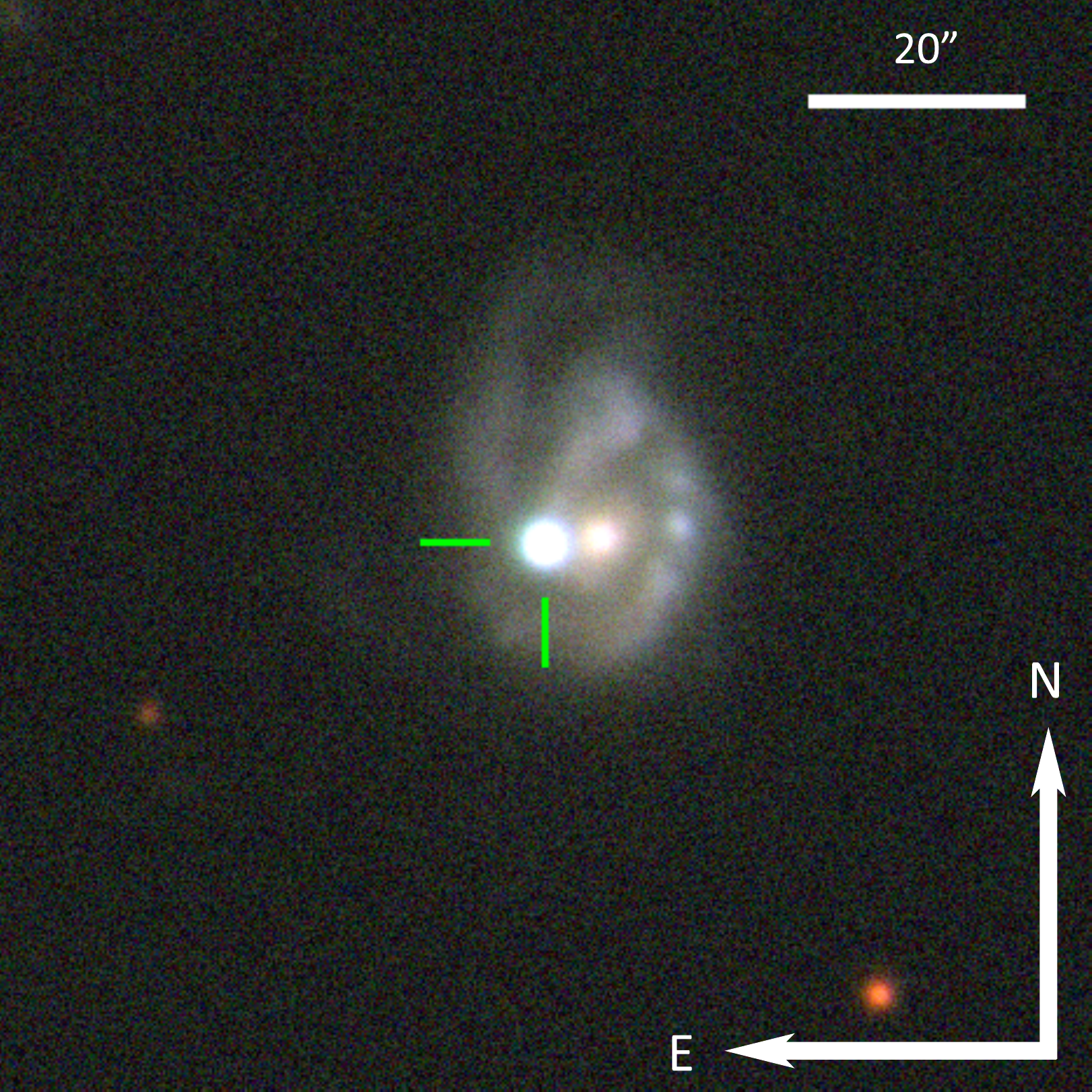}%
	\caption{The \textit{gri} false-color composite image of \host\ with the presence of SN \sn\ observed on July 1, 2017 from NOT+ALFOSC is shown on the right panel. The pre-explosion SDSS image is at left with the SN location marked in a green box.  Both image cutouts are of $100'' \times 100''$ in size.  
	}
	\label{fig:id}
\end{figure*}

The first hydrogen-poor (i.e., Type I) superluminous supernova (SLSN-I), SN 2005ap, was discovered a dozen
years ago by the Texas Supernova Search (TSS), a wide-field, untargeted survey for supernovae (SNe) with a high 
level of spectroscopic completeness \citep{2007ApJ...668L..99Q}.  Subsequent, largely
untargeted, surveys have established the existence of SLSNe-I as a distinct class
of SN (\citealt{2011Natur.474..487Q}).  
SLSNe-I are among the least understood SN populations. The explosion 
mechanism and energy supply mechanism responsible for their extreme peak luminosities ($M_{\rm peak}\lesssim-21$\,mag) and radiated energies are debated \citep{2012Sci...337..927G}, and there are no identifications of progenitor
stars. 

The host of SN 2005ap was a low-metallicity, star-forming, dwarf galaxy, which is
true of almost all subsequent examples of SLSNe \citep[see, e.g.,][]{2011Natur.474..487Q, 2011ApJ...727...15N, 2011ApJ...730...34S, 2013ApJ...763L..28C, 
2014ApJ...787..138L, 2015MNRAS.449..917L, 2016ApJ...830...13P}. There are exceptions such as PTF10uhf \citep{2016ApJ...830...13P} and ASASSN-15lh \citep{2016Sci...351..257D}, which are both in higher mass and metallicity galaxies, although the nature of ASASSN-15lh is debated \citep{2016NatAs...1E...2L, 2017MNRAS.466.1428G}. The underrepresentation of dwarf galaxies in most
galaxy-targeted, professional surveys (e.g., \citealt{2011MNRAS.412.1419L}) and 
in amateur SN searches (see, e.g., \citealt{2017arXiv170402320H})  
would then explain why SLSNe-I were discovered only recently.  This has
also {led the hypothesis} that low
metallicity may be required for the production of SLSNe-I \citep[see, e.g.,][]{2016ApJ...830...13P, 2016arXiv161205978S, 2016arXiv160504925C}.  On the other hand,
the more local SN searches may simply have missed SLSNe-I owing to their
rarity, as their rate is about three orders of magnitude lower than that of
normal Type~Ia SNe \citep{2013MNRAS.431..912Q,2015MNRAS.448.1206M,2017MNRAS.464.3568P}. As an added consequence of their scarcity, few SLSNe-I have been found at sufficiently low redshifts to
permit detailed multiwavelength studies \citep{2016Sci...351..257D, 
2016ApJ...826...39N, 2016ApJ...828....3B,  2017MNRAS.466.1428G, 
2017ApJ...840...57Y, 2017MNRAS.469.1246K}.     

{The explosion mechanism of SLSNe is highly debated and their extreme luminosities cannot be explained as conventional supernovae. Several possible mechanisms have been proposed to explain the powering source, among which the magnetar-spindown \cite[e.g.][]{2010ApJ...717..245K}, pair-instability SNe \citep[PISNe; e.g.,][]{2007Natur.450..390W} and ejecta-CSM interactions  \citep[e.g.][]{2010arXiv1009.4353B,2016ApJ...829...17S} being the most commonly discussed.}
Owing to the perceived preference of SLSNe-I to occur in low-metallicity and dwarf galaxies,{ it has also been proposed \citep{,2014ApJ...787..138L,2016MNRAS.458...84A} that SLSNe-I may be powered by a central engine similar to those in long duration gamma-ray bursts (GRBs), which are also preferentially found in such hosts \citep{2006AcA....56..333S}}.

{Early spectra of SLSNe I show a characteristic ``w"-shaped feature near $ \sim4200 $ \AA\ which is composed of a pair of broad absorption features associated with \Oii\ \citep{2011Natur.474..487Q,2010ApJ...724L..16P}. SLSNe I spectroscopically fall under the classification of type Ic SNe due to absence of any hydrogen, helium and silicon. However, early spectra of SLSNe I are significantly different than the features exhibited by SNe Ic, most notably the w-shaped oxygen feature. The photometric and spectroscopic evolution, and late time energy sources are also significantly different between these two SNe class. However, in many cases their spectra start to show similarities to SNe Ic as the spectra evolves \citep[e.g., SN 2010gx;][]{2010ApJ...724L..16P}.}
     
Here we identify \sn\ (also known as SN 2017egm) as a SLSN-I \citep{2017ATel10498....1D}, and we discuss its discovery and classification in \S2.
The host galaxy,
NGC~3191 (see Figure~\ref{fig:id}), is unusually massive and metal rich, as we discuss
in \S3.
With a redshift $z=0.03063$ \citep{2016arXiv160802013S}, it is the 
closest SLSN-I yet discovered, being a factor of two closer
than the next-nearest example (PTF11hrq at $z=0.057$; \citealt{2016ApJ...830...13P}).
This makes possible the intensive multiwavelength and spectroscopic
observations of this SLSN-I presented in \S4.  We discuss the implications of \sn\ in \S5.
We adopt a luminosity distance of $D_L= 138.7\pm1.9$\,Mpc assuming 
a standard  {\it Planck} cosmology \citep{2016A&A...594A..13P} and foreground Galactic $R_V=3.1$ extinction of 
$E(B-V) = 0.0097\pm0.0005$\,mag \citep{2011ApJ...737..103S}.  The 
blue colors and the absence of narrow \Nai~D absorption 
indicates that there is little additional line-of-sight dust
in the host galaxy.

\section{Discovery and Classification} \label{sec:discovery}

\sn\ ($\alpha=10^{\rm h}19^{\rm m}05.\!\!^{\rm{s}}62$, $\delta=46^{\circ}27'14.\!\!''08$, J2000) was 
discovered by the Photometric Science Alerts Team of the {\it Gaia} mission \citep{2017TNSTR.591....1D} on 
2017 May 23, UT 21:41:13 (JD = 2,457,897.40) at 16.72 mag in the {\it Gaia} $G$ band (UT dates and times are used throughout this paper), and its IAU designation is SN 2017egm.
It was subsequently classified as a Type II SN by \citet{2017ATel10442....1X} 
based on a spectrum taken on 2017 May 26, although \citet{2017ATel10442....1X} 
noted that the object's luminosity ($\sim -19\,$mag) appeared to be 
abnormally bright for a Type II SN. 

In particular,
the source was detected in images taken by the All-Sky Automated Survey for SuperNovae 
(ASAS-SN, \citealt{2014ApJ...788...48S}) starting on 2017 May 20 (JD = 2,457,893.76) at 
$V = 17.36\pm0.14$\,mag (the light-green open-circle $V$-band points in Figure~\ref{fig:lc.app}). 
The ASAS-SN collaboration, working with other groups,  attempts to spectroscopically 
classify all SNe discovered or recovered by ASAS-SN in order to build an unbiased,
nearby SN sample with high spectroscopic completeness \citep[e.g.,][]{2017arXiv170402320H}. 
In this case, the Nordic Optical Telescope (NOT) Unbiased Transient Survey (NUTS) collaboration\footnote{http://csp2.lco.cl/not/} \citep[NUTS;][]{2016ATel.8992....1M} obtained a high signal-to-noise ratio (SNR) NOT/ALFOSC (Andalucia Faint Object Spectrograph and Camera)
spectrum on 2017 May 30.  
This spectrum, as well as a number of subsequent ones (see the top of Fig.~\ref{fig:sp.all}) all showed broad, ``W-shaped"  \Oii\, absorption 
features at rest-frame $\sim4100$~\AA\ and $\sim4400$~\AA\ which are a characteristic of 
most known SLSNe-I \citep{2011Natur.474..487Q}.
This led us to conclude that \sn\ was actually a SLSN-I \citep{2017ATel10498....1D}. Later, \citet{2017arXiv170608517N} duplicated our already public finding.

\section{The host galaxy} \label{sec:host}

\begin{figure}
	\centering
	\includegraphics[width=\linewidth]{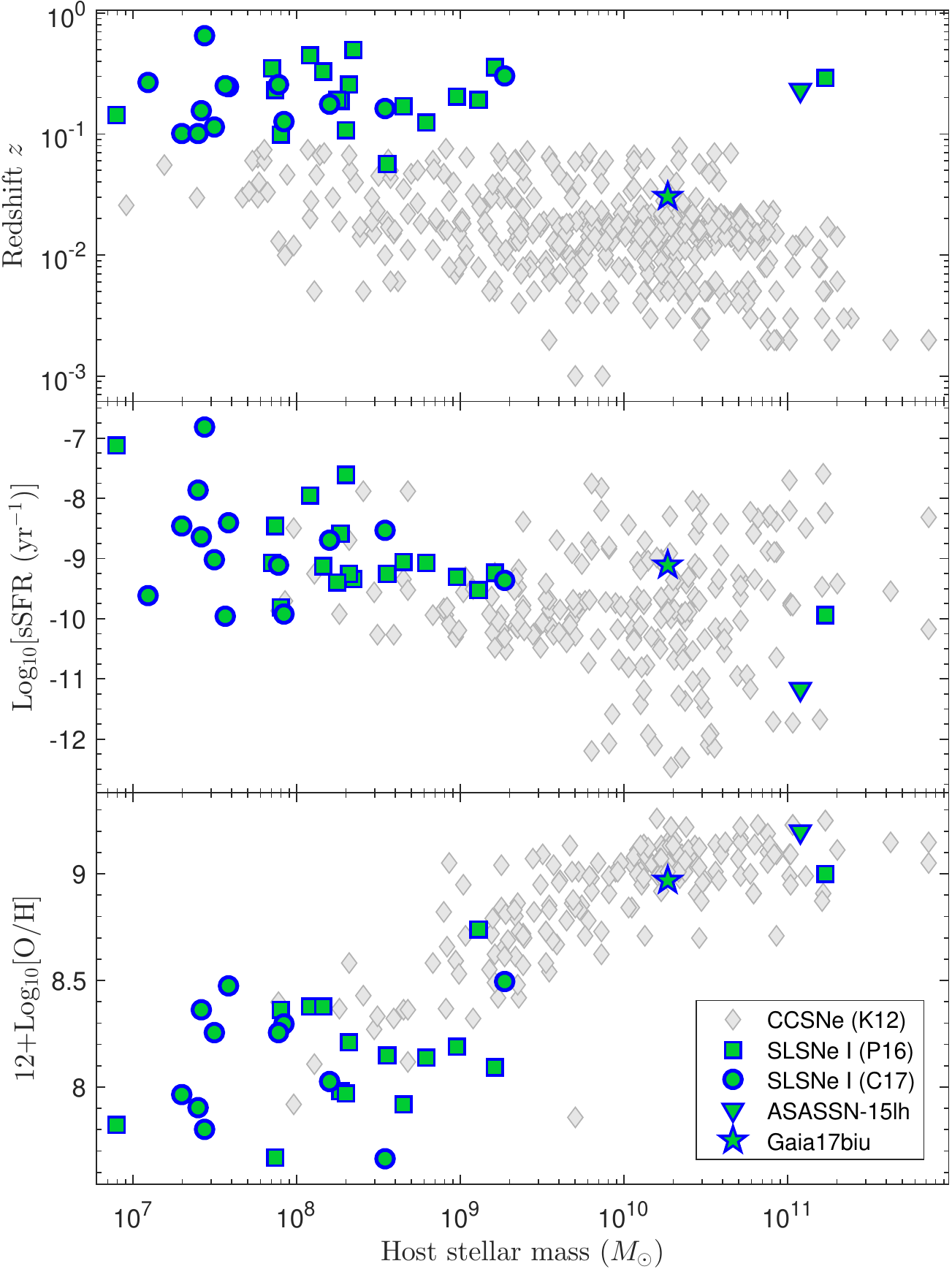}
	\caption{The distribution of SLSNe-I in redshift (top), specific star formation rate (middle),
      and metallicity (bottom) as a function of stellar mass. \sn\ is shown as a green
      star, the green squares and green circles display a comparison SLSN-I sample
      drawn from the PTF \cite{2016ApJ...830...13P} (P16), and the $z<0.3$ sample \cite{2016arXiv160504925C} (C17) augmented by the host of ASASSN-15lh \citep{2016Sci...351..257D, 2017ApJ...836...25M} is shown as green triangle. Some ccSN host-galaxy properties from \cite{2012ApJ...759..107K} (K12) are also shown in grey symbols.
          }
	\label{fig:host_props}
\end{figure}

As pointed out by \citet{2017ATel10498....1D}, the host galaxy of \sn, \host, is atypical for known SLSN-I hosts. It is massive, and it is correspondingly relatively metal rich in accord with the well-established mass-metallicity relation \citep[e.g.][]{2004ApJ...613..898T}.  \cite{2012ApJ...759..107K} analyzed Sloan Digital Sky Survey (SDSS) photometry and spectra of \host\  as the host of the Type II SN PTF10bgl, as well as results from the MPA-JHU analysis of SDSS DR7 galaxies, finding it to be a massive spiral galaxy with a stellar mass of $M_* \approx 5\times10^{10}\,{\rm M}_\sun$ and a central oxygen abundance of 12 + log[O/H] $\approx 8.9$ on the  \cite{2004ApJ...613..898T} strong-line scale. This implies a metallicity of $Z\approx 1.6\,{\rm Z}_\sun$ assuming a Solar oxygen abundance of 12 + log[O/H] $= 8.69$ from \cite{2009ARA&A..47..481A}.  \citet{2017arXiv170608517N} reported properties of the host \host\ based on an archival data analysis, and their results were in agreement with those reported in \citet{2012ApJ...759..107K}.

{
The SDSS spectrum analyzed by \citet{2012ApJ...759..107K} is centered on the
core of the galaxy and offset by $\sim 3$~kpc ($5''$) from the position 
of the SN. In order to estimate the metallicity at the location of the SN, we used the FAST spectrograph on the 60~inch Tillinghast telescope at F. L. Whipple Observatory with a relatively wide slit to obtain a late-time optical spectrum (on 2017 June 21.2)
spanning a wider region of the galaxy and including the SN. After correcting for Milky Way reddening, we find line 
fluxes of  $f(\ha)=9.96\times10^{-14}$\,erg\,s$^{-1}$\,cm$^{-2}$ and $f([\ion{N}{ii}])=3.32\times10^{-14}$\,erg\,s$^{-1}$\,cm$^{-2}$, implying an abundance of 12 + log[O/H]) $= 9.0$ using the \cite{2006A&A...459...85N} 
oxygen abundance calibration for the [\Nii]/\ha\ ratio. This abundance estimator is on the same scale of \cite{2004ApJ...613..898T}.
}

In contrast, most SLSN-I hosts are found in metal poor,
dwarf galaxies \citep{2014ApJ...787..138L}.
\citet{2016ApJ...830...13P} analyzed 32 SLSNe-I discovered
by the Palomar Transient Factory (PTF; \citealt{2009PASP..121.1395L}) and
concluded that they are almost exclusively found in metal-poor and star-forming dwarf 
galaxies having $M_* \lesssim 2\times10^9\,{\rm M}_\sun$ and 12 + log [O/H] $< 8.4$.  
Similar analyses by \citet{2016arXiv161205978S} of 53 SLSNe-I $ z<1  $ and by \citet{2016arXiv160504925C} of SLSNe-I at $z<0.3$ 
concluded that SLSNe-I are strongly suppressed for stellar masses $\gtrsim 10^{10}\,{\rm M}_\sun$, and that SLSN-I production has a metallicity ``cutoff'' at  $\sim0.5\,{\rm Z}_\odot$. 

As a check on the results of \citet{2012ApJ...759..107K} for NGC~3191, we carried out
an independent analysis of its spectral energy distribution (SED).  We fit the SDSS and {\it GALEX} photometry of the host 
using FAST
\citep{2009ApJ...700..221K}, with
the
\cite{2003MNRAS.344.1000B} stellar population synthesis models,
a Chabrier \citep{2003PASP..115..763C} initial mass function (IMF), an exponential star formation history and solar ($ Z=0.02 $) metallicity.  We find
a slightly lower stellar mass of $ \log(M_*/\msun) = 10.21^{+0.17}_{-0.06} $,
 owing to different assumptions about the IMF, and a specific
star formation rate of $\log({\rm sSFR}) = -9.11^{+0.90}_{-0.38}$. This is for an age of $ \rm{log(age)}=8.55^{+0.57}_{-0.20}$ and a star formation timescale of $ \rm{log}(\tau)=8.1^{+0.9}_{-0.1} $.

In Figure~\ref{fig:host_props} we compare the redshift, mass, star-formation rate, and oxygen abundance of \host\ with SLSN-I hosts from the PTF sample by \citet{2016ApJ...830...13P} (P16) and the $z<0.3$ sample by \citet{2016arXiv160504925C} (C17) augmented by the host of ASASSN-15lh ($z=0.2326$; \citealt{2016Sci...351..257D, 2017ApJ...836...25M}). {The oxygen abundance values of C17 are converted from the \cite{2004MNRAS.348L..59P} metallicity scale to \cite{2004ApJ...613..898T} scale using the transformation given by \cite{2008ApJ...681.1183K}} so that all the metallicity estimates are on a common scale. Our methodology for galaxy parameter estimates follows closely those adopted by P16 and C17, so that the comparisons with these samples are made on 
the same stellar mass (using \citealt{2003PASP..115..763C} IMF) and oxygen abundance (using the calibration of \citealt{2006A&A...459...85N}) scales. 
{The sample of hosts from P16 and C17 have stellar masses up to $ 10^9\msun $.
{However, there are few additional SLSNe-I hosts having stellar masses up to $ 10^{10}\msun $ \citep[see, e.g.,][]{2014ApJ...787..138L,2016arXiv161205978S}, but those are not included in the comparison sample (Fig.~\ref{fig:host_props}) due to the lack of oxygen abundance information.}}
The host galaxy of \sn, \host, has a higher mass and metallicity
than the comparison SLSNe-I host sample, although its properties are typical of the general population
of star-forming galaxies (e.g., \citealt{2012MNRAS.422..215Y}) and the hosts
of core-collapse supernovae (ccSNe) (e.g., \citealt{2008ApJ...673..999P, 2012ApJ...759..107K, 2013ApJ...773...12S}).
The only SLSN-I hosts similar to \host\ in mass and metallicity are the host of PTF10uhf in the PTF sample and ASASSN-15lh \citep{2016Sci...351..257D, 2017ApJ...836...25M}. 
Since the redshift of PTF10uhf is typical of the other SLSNe-I in the PTF sample, the rarity of additional
higher mass and metallicity hosts drives the conclusion that low metallicity is 
favored for producing SLSNe-I. 
{However, \sn\ is found at a record-breaking low redshift, and the relative deficiency of low-redshift {($ z<0.05 $)} SLSNe-I with dwarf hosts implies that any suppression of SLSN-I production in metal-rich and massive hosts is likely weaker than previously thought.}

{The location of \sn\ has an offset of $ 5\farcs16 $ (3.47 kpc) from the center of \host, which after normalizing by the half-light radius (\textit{r}-band), implies an offset of 0.67. This is somewhat on the lower side as compared to the distribution found for SLSN-I hosts \citep[e.g.][]{2015ApJ...804...90L,2016arXiv160504925C} having a median normalized offset of $ \sim1 $. Interestingly, SLSNe with massive hosts in these samples tend to have larger offsets, which is opposite to that observed in the case of \sn.} 

\section{Optical observations} \label{sec:opt_observation}
\subsection{Data collection and reduction}
In addition to the ASAS-SN \textit{V}-band observations, multiband optical photometric observations were obtained with the Apogee Alta U230 camera at Post Observatory SRO (CA, USA) and the Apogee Alta U47 at Post Observatory Mayhill (NM, USA) with 0.6~m telescopes at both locations, the
0.5~m DEdicated MONitor of EXotransits and Transients \citep[DEMONEXT;][]{2016SPIE.9906E..2LV} and the 0.5~m Iowa Robotic Telescope (both at the Winer Observatory, AZ, USA), the IO:O imager on the 2.0~m Liverpool Telescope (LT) at La Palma, ALFOSC and NOTCam on the 2.5~m NOT at La Palma, the Las Cumbres Observatory 1.0~m telescope network \citep{2013PASP..125.1031B},  the 1.0~m Nickel telescope at Lick Observatory (CA, USA), the 1.0~m telescope at Weihai Observatory of Shandong University (China)   \citep{2014RAA....14..719H}, the 2.0~m Ritchey-Chretien telescope at Bulgarian National Astronomical Observatory (Rozhen, Bulgaria), and the Meade 10~inch LX-200 Schmidt-Cassegrain Telescope at Antelope Hills Observatory (CO, USA).  

We triggered observations with \textit{Swift} \citep{2004ApJ...611.1005G} lasting from 2017-06-02 to 2017-07-04 (PI S. Dong, \textit{Swift} Target IDs 10150 and 10154) to obtain near-UV (NUV) observations with the Ultraviolet Optical Telescope (UVOT) \citep{2005SSRv..120...95R}.
Except for the ASAS-SN difference imaging analysis pipeline, point-spread-function (PSF) photometry was done with the DoPHOT \citep{1993PASP..105.1342S} package for the optical and near-infrared (NIR) data.
Optical photometric data were calibrated using APASS\footnote{https://www.aavso.org/apass} standards  for the Johnson $BV$ filters and the Sloan (AB magnitude system) $gri$ filters, and the \textit{JHK} NIR data were calibrated using 2MASS \citep{2006AJ....131.1163S} standards. 
The \textit{Swift}/UVOT photometry was measured with UVOTSOURCE task in the Heasoft package using $5''$ apertures and placed in the Vega magnitude system, 
adopting the revised zero-points and sensitivity from \cite{2011AIPC.1358..373B}. The reduced photometry is reported in Table~2.

Optical spectroscopic observations were obtained using ALFOSC on NOT,
the Kast Dual Channel Spectrograph mounted on the 3~m Shane telescope at Lick Observatory \citep[CA, USA][]{1993lick...miller}, the B\&C spectrograph mounted on the 1.2~m Galileo Telescope and the AFOSC spectrograph on 1.8~m Copernico telescope in Asiago (Italy), the FAST spectrograph \citep{1998PASP..110...79F} mounted on 60'' Tillinghast telescope at F. L. Whipple Observatory 
(AZ, USA), and the SPRAT spectrograph mounted on 2.0~m Liverpool telescope in La Palma. 
Most spectra were taken at or near the parallactic angle \citep{1982PASP...94..715F} to minimize differential slit losses caused by atmospheric dispersion.  The log of optical spectroscopic observations is given in Table~\ref{tab:speclog}.

The ALFOSC and AFOSC data were reduced using \textsc{FOSCGUI}\footnote{Developed by E. Cappellaro; http://sngroup.oapd.inaf.it/foscgui.html}.  The FAST spectra were reduced with the standard pipeline system using IRAF\footnote{IRAF is distributed by the National Optical Astronomy Observatory, which is operated by the Association of Universities for Research in Astronomy (AURA) under a cooperative agreement with the US National Science Foundation.} scripts developed for FAST and Massey standards \citep{1988ApJ...328..315M, 1990ApJ...358..344M}  for spectrophotometric calibration.
SPRAT spectra were reduced and flux calibrated using the LT pipeline \citep{2012AN....333..101B,2014SPIE.9147E..8HP}. 
Kast data were reduced following standard techniques for CCD processing and spectrum extraction \citep{2012MNRAS.425.1789S} utilizing IRAF routines and custom Python and IDL codes\footnote{https://github.com/ishivvers/TheKastShiv}. Low-order polynomial fits to comparison-lamp spectra were used to calibrate the wavelength scale, and small adjustments derived from night-sky lines in the target frames were applied.  Observations of appropriate spectrophotometric standard stars were used to flux calibrate the spectra. 
Spectra from other instruments were  reduced and calibrated using standard
procedures. Telluric corrections were applied to remove strong atmospheric absorption bands. For some spectra where appropriate telluric standards were not available, we manually remove the region strongly affected by telluric features. 
We also obtained spectra of \sn\ in the NIR at $-3$~days using the Aerospace Corporation's Visible and Near-Infrared Imaging Spectrograph (VNIRIS) on the Lick Observatory 3~m Shane reflector, near maximum light ($-0.5$~days) using the Spex medium-resolution spectrograph \citep[0.7--5.3\,$\mu$m;][]{2003PASP..115..362R} on the NASA Infrared telescope facility (IRTF), and at $-0.8$~days with NOTCam. The spectra are reduced and calibrated using standard methods.

Three epochs of spectropolarimetry were obtained using the polarimetry mode of Lick/Kast on 2017 June 21 ($-1$~day), June 27 ($+5$~day), and July 1 ($+9$~day). On June 21 the source was observed just after $12^\circ$ twilight and over a moderately high airmass range of 1.58--1.89, as it was setting. The June 26 and July 1 data were obtained progressively deeper into twilight, as the sky position of the setting source became increasingly unfavorable. Polarization spectra were measured at each of four waveplate angles ($0^\circ$, $45^\circ$, $22.5^\circ$, and $67.5^\circ$), with three exposures obtained at each angle to remove cosmic rays via median combination. The individual exposures were 270\,s, 200\,s, and 140\,s for the June 21, 26, and July 1 observations (respectively), for total integration times of 3240\,s, 2400\,s, and 1680\,s over all waveplate angles. Low-polarization standard stars were observed to calibrate the instrumental position-angle curve with respect to wavelength, and to confirm that the instrumental polarization was negligible. High-polarization standard stars were observed to calibrate the polarization position angle on the sky, $\theta$. All of the spectropolarimetric reductions and calculations follow the methodology described by \cite{2014MNRAS.442.1166M,2015MNRAS.453.4467M}, and the polarimetric parameters are defined in the same manner. We refer the reader to those works for more detailed information on the observations and reductions.

\begin{figure}
	\centering
	\includegraphics[width=\linewidth]{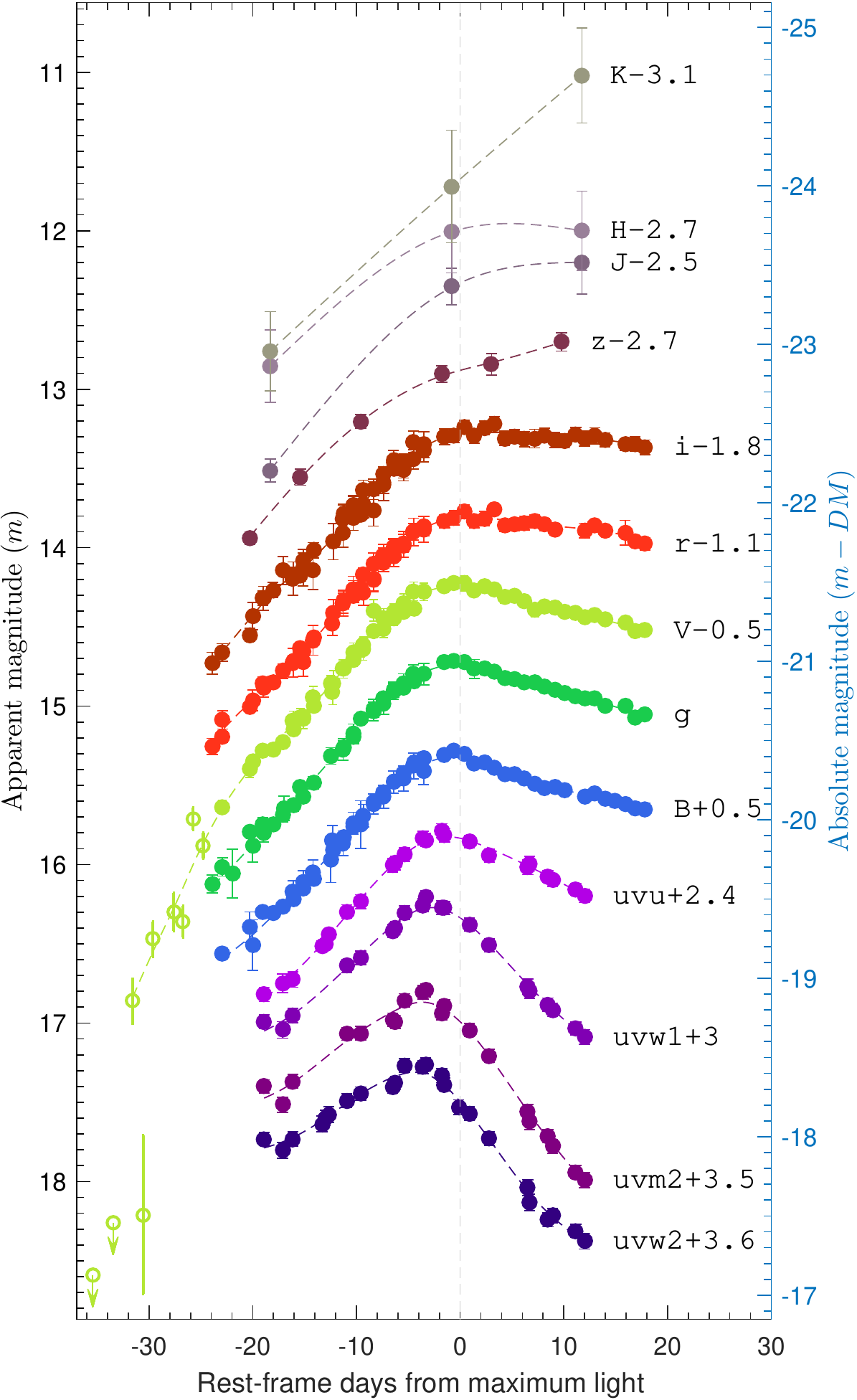}
	\caption{The Johnson-Cousins \textit{BV}, SDSS \textit{griz}, \textit{Swift}-UVOT NUV (Vega magnitude), and $JHK$  light curves
		of \sn.  The light curves are shifted vertically for clarity. 
		The reference epoch is set by the \textit{g}-band maximum (\EpEpoch).
		Low order splines are shown to connect the data for visual clarity.
		{The open circles (light-green) are \textit{V}-band detections from ASAS-SN, and the open circles with downward arrows represent ASAS-SN upper limits.}}
	\label{fig:lc.app}
\end{figure}

\subsection{Photometric evolution} \label{sec:lc}

In Figure~\ref{fig:lc.app} we show the full set of multiband light curves, where we adopt the peak\footnote{The peak was found by fitting a fourth-order polynomial to the flux values close to maximum brightness ($\pm15$ days).} of the
$g$-band light curve at $\rm{\EpEpoch\pm 0.7} $ (21.8 June 2017) as our reference epoch throughout this paper. 

\begin{figure}
	\centering
	\includegraphics[width=\linewidth]{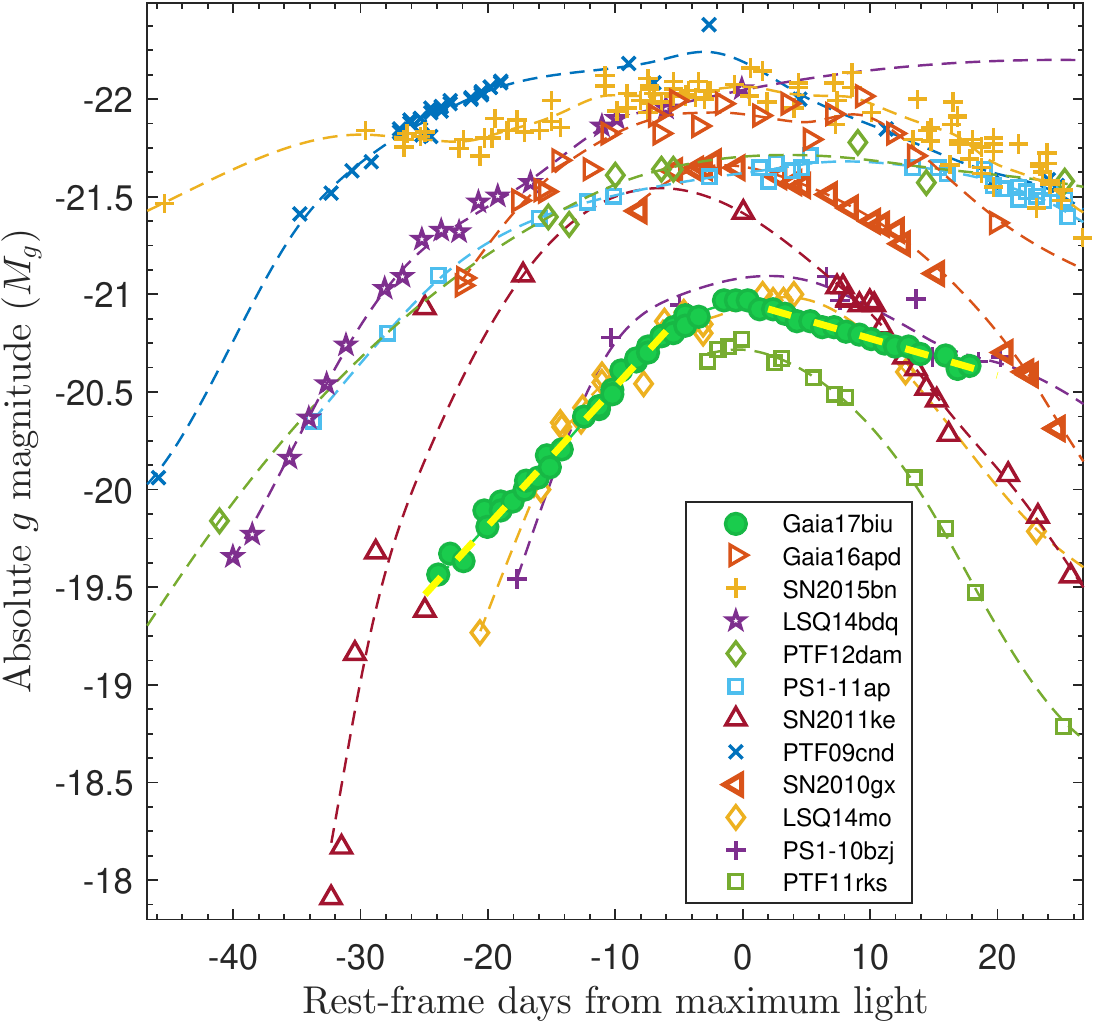}%
	\caption{
		The absolute $g$-band light curve of \sn\ as compared to other SLSNe-I. The sample is mainly based on that in \citet{2015MNRAS.452.3869N} with the additions of two recently discovered low-$z$ SLSNe-I, SN 2015bn and Gaia16apd. The full sample includes Gaia16apd \citep{2017MNRAS.469.1246K}, SN 2015bn \citep{2016ApJ...826...39N}, LSQ14bdq \citep{2015ApJ...807L..18N}, PTF12dam \citep{2013Natur.502..346N}, PS1-11ap \citep{2014MNRAS.437..656M}, SN 2011ke \citep{2013ApJ...770..128I}, PTF09cnd \citep{2011Natur.474..487Q}, SN 2010gx \citep{2010ApJ...724L..16P}, LSQ14mo \citep{2017A&A...602A...9C}, PS1-10bzj \citep{2013ApJ...771...97L}, and PTF11rks \citep{2013ApJ...770..128I}. 
		A pair of straight lines (yellow dashed) are shown on the rising and declining parts of \sn\ to illustrate their remarkable linearity.
		}
	\label{fig:lc.abs}
\end{figure}

After correcting for Galactic extinction and applying small K-corrections based on the optical spectroscopy, \sn\ peaked at $M_g=-20.97\pm0.05$\,mag, 
{which is close to the mean SLSNe-I peak magnitude \citep{2017arXiv170801623D,2017arXiv170801619L}}.
In Figure~\ref{fig:lc.abs} we compare its $g$-band light curve to those of a number  of other SLSNe-I mainly from the sample of \citet{2015MNRAS.452.3869N}  and adding SN~2015bn \citep{2016ApJ...826...39N} and Gaia16apd \citep{2017MNRAS.469.1246K}. 

Almost all well-observed SLSNe-I appear to show significant curvatures in their light curves near their peaks (see Figure~\ref{fig:lc.abs}). Some SLSNe-I (the most conspicuous example is SN 2015bn \citealt{2016ApJ...826...39N}, but also LSQ14bdq and LSQ14mo) show sporadic undulations in their light curves. In contrast, the light-curve evolution of \sn\ from $\sim -20\,{\rm day}$ to $\sim 20\,{\rm day}$ can be almost perfectly described by a linear rise followed by a linear decline in magnitude (shown as yellow dashed lines in Figure~\ref{fig:lc.abs}) with a rapid (few-day) turnaround at the peak. The linearity of the light curves implies that the SN luminosity is evolving exponentially in both the rise to the peak and decline from the peak. Such photometric evolution appears to be unprecedented among SLSNe-I. 
The rise time for \sn, characterized by the $e$-folding time $\tau_{g-rise} \approx 20$~days before
the peak, is relatively fast.  {This is consistent with the general trend that the less
luminous SLSNe-I have faster rise times, as evident from Figure~\ref{fig:lc.abs}.
\cite{2014ApJ...796...87I} also noted a similar positive correlation, but between the decline timescale and the luminosity. However, \sn\ is one of the slowest declining SLSN-I despite having a lower luminosity. A few other SLSNe have also been found to not follow this correlation, most notably PS1-14bj \citep{2016ApJ...831..144L} which has both long rise and decline timescales.}
We have insufficient observational coverage to directly measure one $e$-folding in flux in the decline of \sn. Nevertheless, following the almost perfect exponential flux decline as we see here, the $e$-folding decline time is estimated to be $\tau_{g-decl.} \approx 60$ days. 
Thus,  \sn\ combines one of the fastest rise times with one of the slowest declines.

\citet{2016MNRAS.457L..79N} studied light curves of several SLSNe-I showing possible double peaks similar to LSQ14bdq \citep{2015ApJ...807L..18N} and proposed that an early-time luminosity excess is ubiquitous in SLSNe-I. However, we do not see clear evidence for such an early, pre-peak bump in the smoothly evolving ASAS-SN \textit{V}-band light curve of \sn. There is a possible dip with $V = 18.7\pm 0.5$\,mag at 2017-05-21.30 (phase $= -30.6$\,days), as compared with $V=17.4\pm0.1$\,mag one day earlier ($-31.6$\,day) and $V=17.0\pm 0.1$\,mag one day later ($-29.6$\,days), but the evidence for a dip has low statistical significance and the implied timescale ($\sim 1\,{\rm day}$) is considerably shorter than the early bump seen in LSQ14bdq \citep{2015ApJ...807L..18N}. {Overall, there is no evidence in our data for any significant departure (including precursor ``bumps'' as reported in \citealt{2016MNRAS.457L..79N} or ``undulations'' as seen in the light curves of SN 2015bn by \citealt{2016ApJ...826...39N}) from a smooth light-curve evolution.}

\begin{figure}
	\centering
	\includegraphics[width=8.75cm]{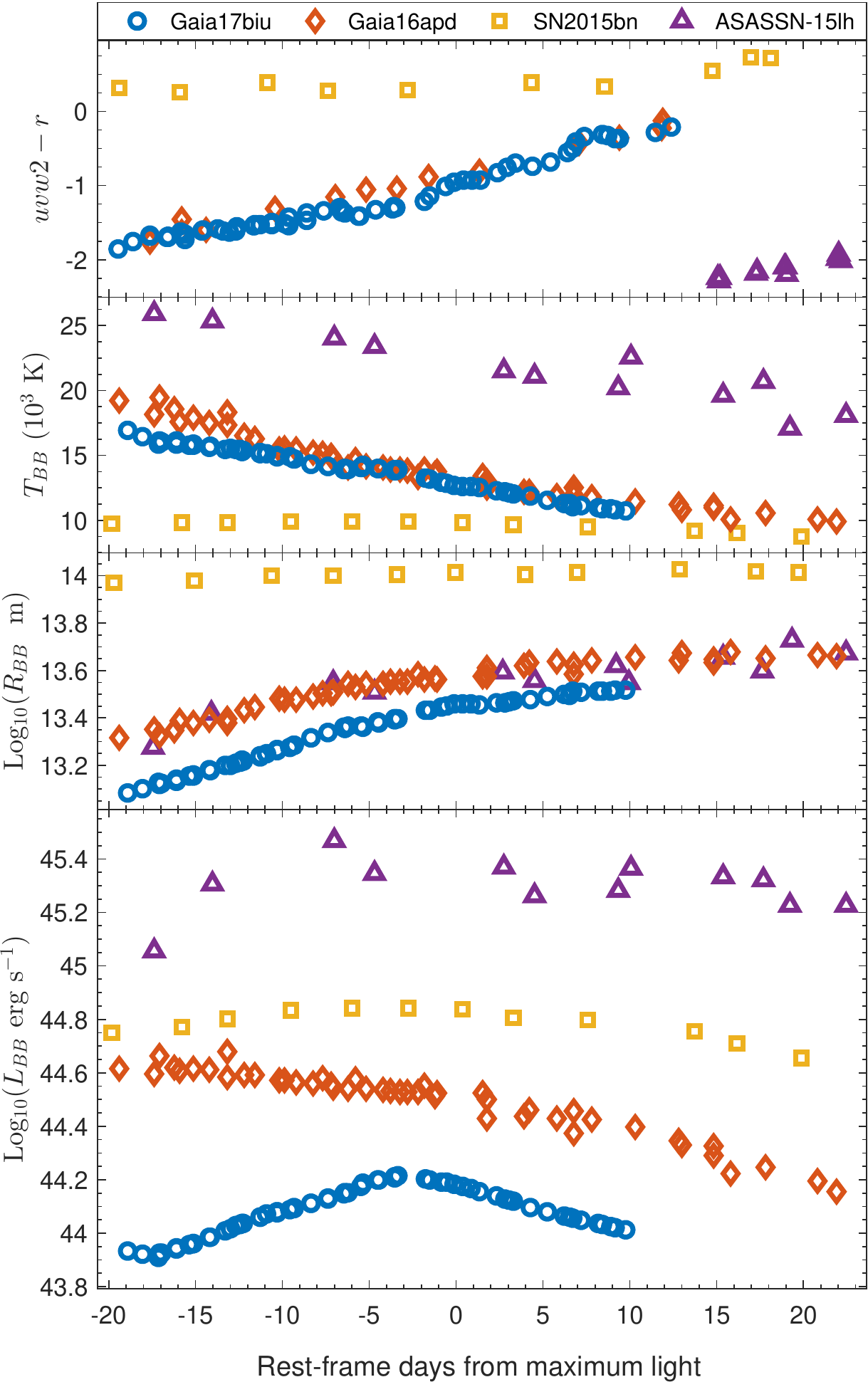}
	\caption{The evolution of NUV (\textit{uvw2}) to optical (\textit{r}) colors, black-body temperature, radius, and luminosity of \sn\ as compared to the handful of SLSNe-I having good NUV and optical coverage.}
	\label{fig:color_TRL}
\end{figure}

Only a handful of low-redshift SLSNe-I have similar wavelength coverage with good cadence like that for \sn.  
Figure~\ref{fig:color_TRL}~[Panel~1] shows the evolution of the {\it Swift} NUV (\textit{uvw2} band at 2080\,\AA) to optical (SDSS \textit{r} band at 6254\,\AA)
color of \sn\ and three other SLSNe-I
where such data are available (Gaia16apd, \citealt{2017MNRAS.469.1246K}; SN 2015bn, \citealt{2016ApJ...826...39N}; ASASSN-15lh, \citealt{2016Sci...351..257D}). 
Due to lack of NUV spectroscopic observations for (most phases of) these comparison SNe, we assumed black-body SEDs to compute and apply K-corrections based on effective temperatures.
\sn\ closely follows the color evolution of Gaia16apd, which has one of
the bluest UV to optical colors among SLSNe-I \citep{2017MNRAS.469.1246K, 2017ApJ...840...57Y}.
\citet{2017ApJ...840...57Y} attribute this blue color to reduced line blanketing
due to both the newly synthesized metals in the ejecta and the likely sub-solar progenitor as deduced from its dwarf host. The latter reasoning is difficult to apply    to \sn\ owing to its relatively high host-galaxy metallicity (see \S3).

We also fit the NUV through $z$-band photometry of \sn\ with black-body
SEDs\footnote{The SED is redshifted to the observed frame prior to filter-response convolution and fitting.}. Figure~\ref{fig:color_TRL}~[Panels~2-4] shows the
resulting rest frame estimates for the evolution of the effective temperature, black-body
radius, and bolometric luminosity.  The black-body models fit the SEDs
well.  \sn\ evolves in temperature like Gaia16apd but has a significantly
smaller photosphere and hence luminosity.  Gaia16apd evolves in radius 
like ASASSN-15lh but is significantly cooler and hence less luminous.  
SN~2015bn is cooler, but larger in radius, than \sn, Gaia16apd, and ASASSN-15lh,
leading to a luminosity intermediate to those of the other three sources.  While
the sample of SLSNe-I with good multiwavelength photometry is limited,
the population appears to show a considerable diversity in size and
temperature to accompany the range of luminosities.
{We note that near $ -19 $d there is an apparent short decline lasting for only two days. However, we do not consider it to be a significant indication for a ``bump", as this originated from only one epoch (-18.9d) of data point in UVOT-NUV bands (see Fig.~\ref{fig:lc.app}). Even though we have significant optical observations before -19d, but we do not find any such indication of a bump.}

\begin{figure*}
	\centering
	\includegraphics[width=0.76\linewidth]{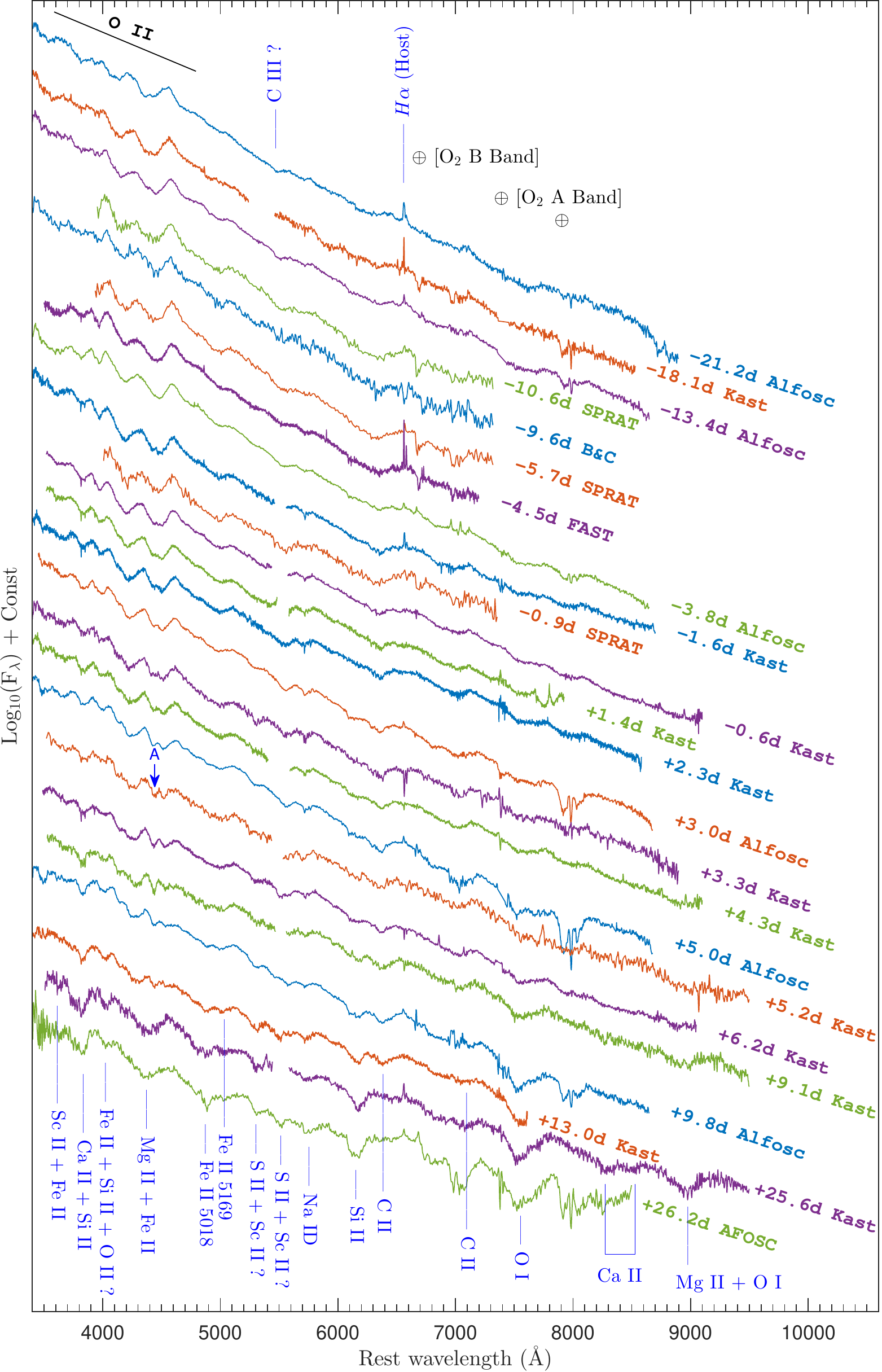}
	\caption{
		Rest-frame spectral evolution of \sn. The wavelength range for the features attributed to O~II is indicated by the black line at top. Other spectral features are marked by vertical blue lines. The arrow marked ``A'' indicates an unidentified spectral feature previously not detected in SLSNe-I (discussed in \S4.2). Each spectrum is labeled by the instrument used and the rest-frame phase from maximum light.}
	\label{fig:sp.all}
\end{figure*}

\subsection{Spectroscopy} \label{sec:sp}

The spectroscopic evolution of \sn\ is shown in Figure~\ref{fig:sp.all}. As already noted in \S2, it exhibits the strong W-shaped \Oii\ absorption lines at rest-frame $\sim4100$\,\AA\ and $\sim4400$\,\AA\ that are characteristic of most known SLSNe-I. Our earliest spectra show these features at $\sim 20,000$\,\kms\, with broad, extended, and somewhat flat-bottomed shapes. As the velocities decrease with time, the line shapes become sharper and more similar to those exhibited by SN 2010gx \citep{2010ApJ...724L..16P} (see the comparison in Fig.~\ref{fig:sp.speccomp}). After reaching peak brightness, the \Oii\ features start to become weaker and are overtaken by other metallic lines.  

\begin{figure*}
	\centering
	\includegraphics[width=0.9\linewidth]{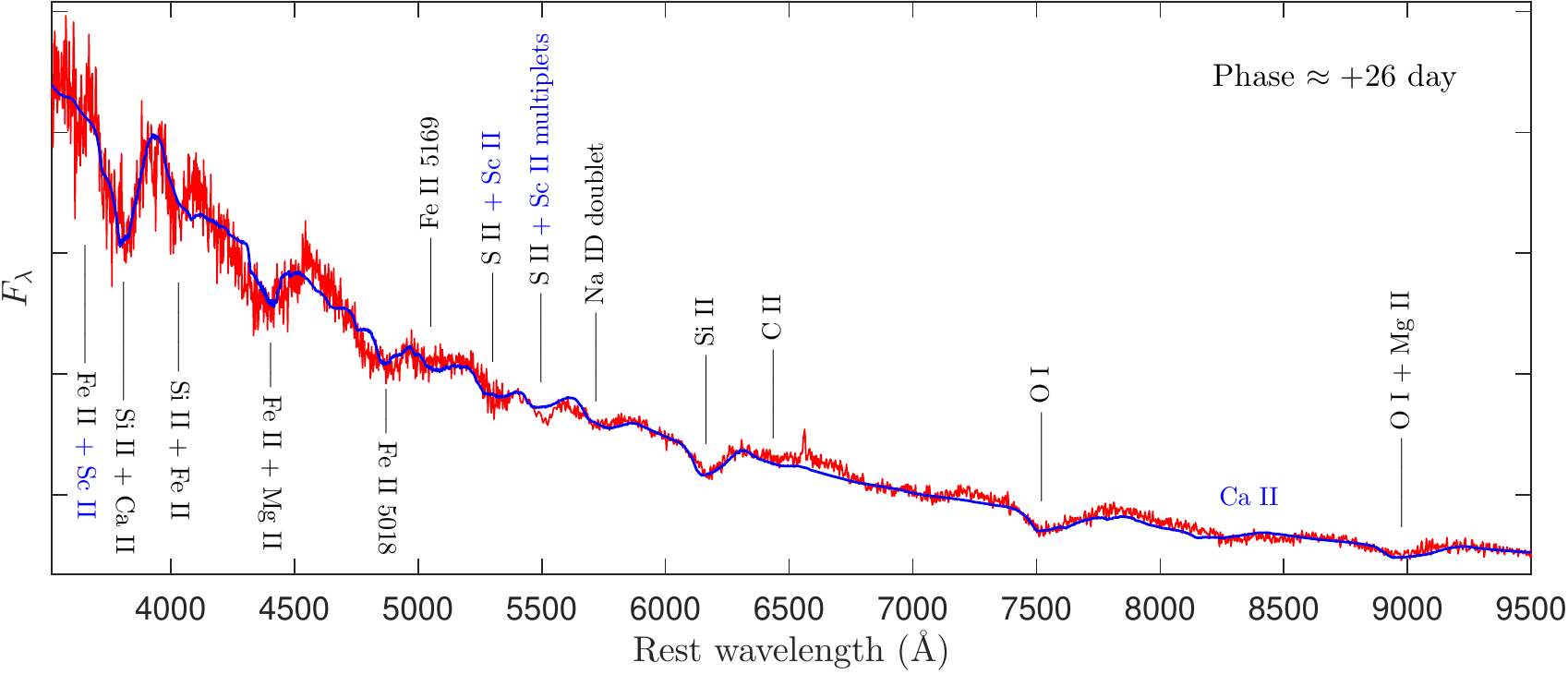}%
	\caption{\synow\ model spectrum (blue solid line) is generated to reproduce the observed $\approx +26 $ day \sn\ spectrum (red solid line). The line velocity for all the identified features is $ \approx10,500 $~\kms. The ions labeled in black are used to synthesize the corresponding spectral features. The ions labeled in blue are not used in the model, but are identified based on their wavelengths and assuming the same velocity as for the lines used in the model.
	}
	\label{fig:synow}
\end{figure*}

{To identify the spectral features in \sn\ we model the spectra using  \synow\ \citep{1997ApJ...481L..89F,1999MNRAS.304...67F,2002ApJ...566.1005B}. \synow\ is a parameterized spectrum synthesis code with an underlying LTE continuum, assuming pure resonant scattering and radiative transfer is simplified using Sobolev approximation. We selected to model the latest spectra for having the most prominent spectral features. The $ +25.6 $ day Kast spectrum is used with the missing portion near $ 5500 $\AA\ stitched with $ +26.2 $ day AFOSC spectrum. In Fig.~\ref{fig:synow} we show the best fit model spectrum using a combination of \Oi, \Feii, \Nai, \ion{S}{ii}, \ion{C}{ii}, \ion{Mg}{ii}, \Siii\ and \Caii\ atomic species. An exponential optical depth profile is found to be suitable for reproducing the observed line profiles. All the spectral features are formed at a single velocity of roughly $\approx 10,500\pm1,000 $ \kms, which further confirms our line identifications. In Figure 6, the ions labeled in black are used in \synow\ to synthesize the corresponding spectral feature in the model spectrum. The ions labeled in blue are identified based only on their velocity, but has not been used to produce the corresponding model feature. These ions can also reproduce P-Cygni profiles at the labeled location using the exact same velocity as for other lines (i.e $\approx 10,500 $ \kms). However, at the same time the given ion will also produce several additional set of features (of relative strengths) in the model which are not present in observed spectrum. 
The \Scii/\ion{S}{ii} features near 5500\AA\ are prominent examples of such an ambiguous identification. The model spectrum can produce these features using \ion{S}{ii} as well as \Scii\ at similar wavelengths and using same velocity. However, increasing the strength of \Scii\ to match the target feature would also produce unmatched strong features near 4100\AA. On the other hand, if the progenitors of SLSNe are massive, the presence of \Scii\ is more likely than \ion{S}{ii}. The ambiguity of \Scii\ (and other features labeled in blue) can also be due to unknown complexity of radiation transfer and non-LTE SED for \sn, which are beyond the simplified assumptions in \synow. We also modeled the $ +13 $~day spectrum to confirm the identification of \ion{C}{ii} features near 6400\AA\ and 7100\AA\ (as labeled in Fig.~\ref{fig:sp.all}). }

At $\sim 4450$\,\AA, close to the longer wavelength doublet component of the W-shaped \Oii\ feature, we find an apparent double absorption feature that is 
labeled as ``A'' in Figure~\ref{fig:sp.all}.  Such a feature has not previously
been observed in SLSNe-I to our knowledge.  It is most prominent in the $\sim 5\,{\rm day}$ spectrum, and still appears to be present but with a different shape in the $\sim 10\,{\rm day}$ spectrum. The feature can be traced back to the earliest spectrum, where it is likely weakly blended with the longer wavelength doublet component of the W-shaped feature, possibly (partly) contributing to its flat-bottomed profile. As the ejecta velocities decline, this feature becomes more clearly resolved. 

A shallow feature is also visible in the early-time spectra near 5500\,\AA\ until the $\sim +5\,{\rm day}$  spectrum. This feature is also visible in SN~2015bn and Gaia16apd, and has been attributed to \ion{C}{iii} $\lambda$5690. Another 
broad feature near 6300\,\AA\ is attributed to \ion{C}{ii} $\lambda$6580 \citep{2017ApJ...840...57Y}, {which we also find in our \synow\ modeling of the later phase spectra}. One prominent metallic line is \Feii\ $\lambda$5169. This feature appeared at $\sim -13\,{\rm day}$ with $\sim 10,000$\,\kms\ and remained until the last spectrum with little velocity evolution. Such an non-evolving \Feii\ line has been also been observed in other SLSNe-I \citep[e.g.,][]{2017MNRAS.469.1246K,2015MNRAS.452.3869N}. 

Apart from these features, the early-time spectra are mostly devoid of other prominent features, while at later phases, heavily blended metallic lines start to appear. Blends of a few {\Feii, \Nai~D, and possible \ion{S}{ii}/\Scii\ multiplets near $ 4900-5600 $\,\AA\ }can be identified in the $ +13 $~day spectrum and became more prominent at later phases. Such a clear detection of  these \Feii\ and {\ion{S}{ii}/\Scii\ }metallic lines in SLSNe is uncommon. In particular, the $\sim +26$ day spectrum shows a remarkable transition compared to the previous spectra: later spectra are dominated by numerous strong metal-rich features. At this phase, we see the emergence of the \Caii\ $\lambda\lambda$3969, 3750 doublet along with \Feii\ $\lambda$5018, the \Nai~D $\lambda\lambda$5890, 5896 doublet, and the \Caii\ $\lambda\lambda$8498, 8542, 8662 NIR triplet. Other features which became significantly stronger than in previous spectra are \Siii\ $\lambda$6355, \Oi\ $\lambda$7774, and \ion{Mg}{ii} near 9000\,\AA. This likely marks the start of the transition to the spectrum of a normal SN~Ic, as seen in some other SLSNe-I \citep{2010ApJ...724L..16P}.

 \begin{figure}
	\centering
	\includegraphics[width=\linewidth]{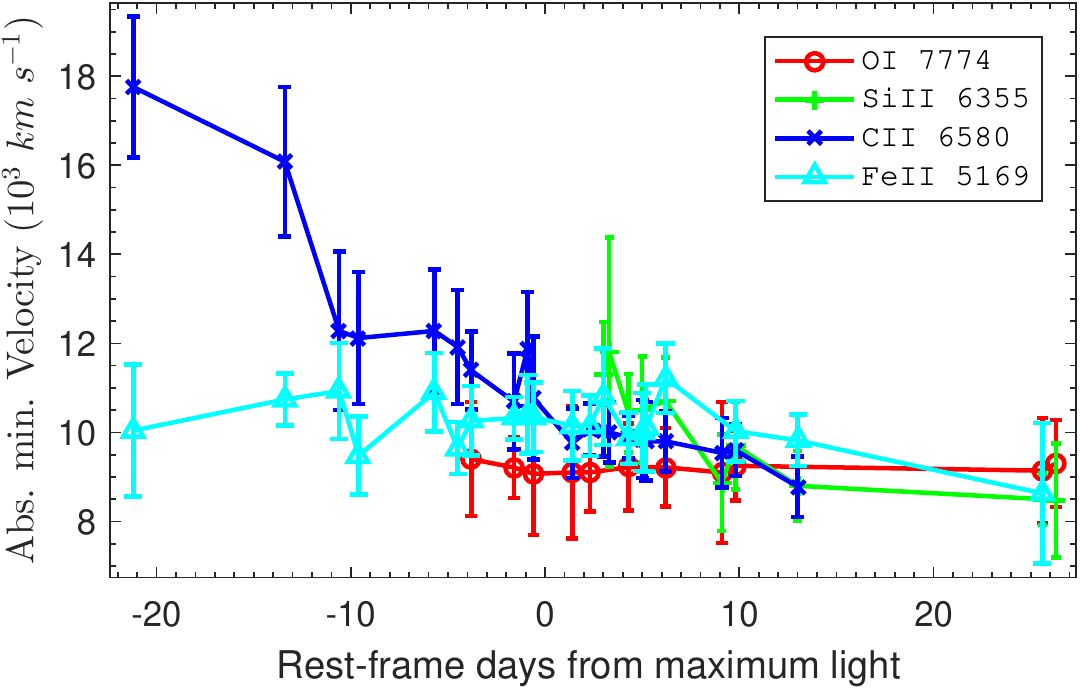}%
	\caption{The line velocity evolution for \Feii\ \ld5169, \Siii\ \ld6355, \Oi\ \ld7774 and \ion{C}{ii} \ld6580. The velocities are estimated from the absorption minima of the corresponding P-Cygni profiles.
	}
	\label{fig:vel_prof}
\end{figure}
 \begin{figure}
	\centering
	\includegraphics[width=\linewidth]{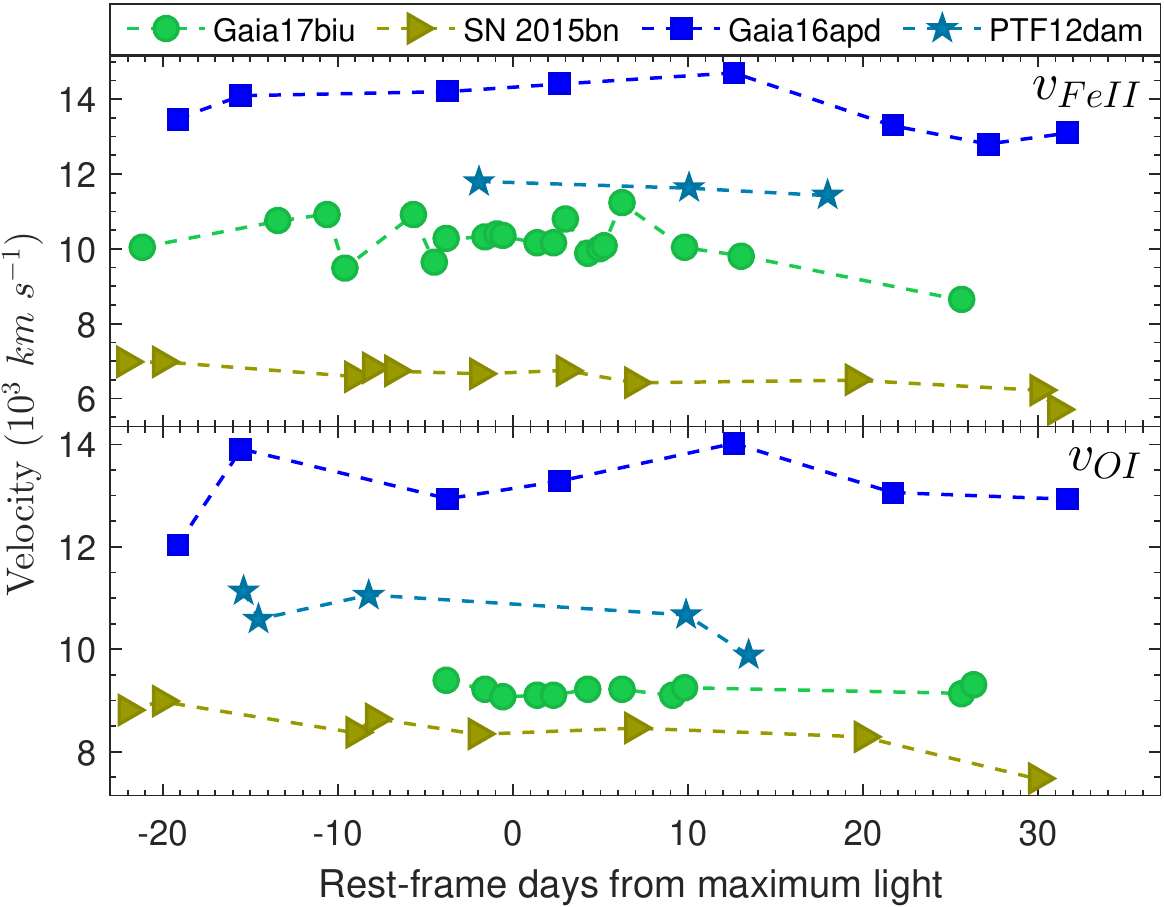}%
	\caption{The \Feii\ and \Oi\ \ld7774 line velocities of \sn\ as compared to SLSNe 2015bn \citep{2016ApJ...826...39N}, Gaia16apd \citep{2017MNRAS.469.1246K} and PTF12dam \citep{2013Natur.502..346N}. The \Feii\ velocities are measured using the \ld5169 line, except for SN 2015bn, where \ld4924 line is used. In case of PTF12dam \Feii\ \ld5169 velocities are presented in \cite{2015MNRAS.452.3869N} while \Oi\ velocities are measured from spectra \citep{2013Natur.502..346N}.
	}
	\label{fig:vel_comp}
\end{figure}

{Figure~\ref{fig:vel_prof} shows the velocity evolution of the \Feii\ \ld5169, \Siii\ \ld6355, \Oi\ \ld7774 and \ion{C}{ii} \ld6580 lines, estimated by measuring the absorption minima of the P-Cygni profiles. These lines are well-identified and free from strong blending with neighboring lines in all the spectra. The line velocities, particularly for \Feii\ and \Oi, remains almost constant with time. This is commonly observed in other SLSNe-I as well \citep[e.g.,][see also Fig.~\ref{fig:vel_comp}]{2015MNRAS.452.3869N,2016ApJ...826...39N}. The constant velocity evolution may possibly indicate stratification of line forming shells within a homologous expansion. On the other hand, the \ion{C}{ii} velocity shows a monotonic decline with time, which is consistent with a typical spherical-SN model where deeper and slower moving layers are exposed by a receding photosphere. In Figure~\ref{fig:vel_comp} we compare the \Feii\ \ld5169 and \Oi\ \ld7774 velocity evolution with other well observed SLSNe-I 2015bn \citep{2016ApJ...826...39N}, Gaia16apd \citep{2017MNRAS.469.1246K} and PTF12dam \citep{2013Natur.502..346N}. For SN 2015bn, the \Feii\ \ld4924 line velocity is used because the \Feii\ \ld5169 line profile appears to be contaminated by \ion{Fe}{iii} emission in the early phases. \sn\ and the comparison sample all show very little velocity evolution. The \Feii\ line velocity for \sn\ remains almost constant at $ \sim10,000 \kms$, which is very close to the median velocity of $ 10,500 \kms$ for the SLSNe-I sample complied by \cite{2015MNRAS.452.3869N}.}

Since \sn\ is the apparently brightest SLSN observed to date by a factor of nearly 10, it provided an unprecedented opportunity to obtain high-SNR spectra. In Figure~\ref{fig:sp.speccomp} we have marked 
several additional broad or weak spectroscopic features that apparently have not been previously seen in any SLSN-I spectra, presumably because of their typically lower 
SNRs. We also note that some of these features appear to be only visible for short periods of time, and possibly our high spectroscopic cadence has helped in capturing \sn\ during such transitions.
These features could be blended metallic lines that become more visible as
the line velocities decrease. A few of these features near 5400\,\AA\ are likely associated with metallic lines such as \Feii\ and {\ion{S}{ii}/\Scii,} which become more prominent at later phases (see the $\sim 25$\,day spectra; Fig.~\ref{fig:sp.all}).

\begin{figure}
	\centering
	\includegraphics[width=\linewidth]{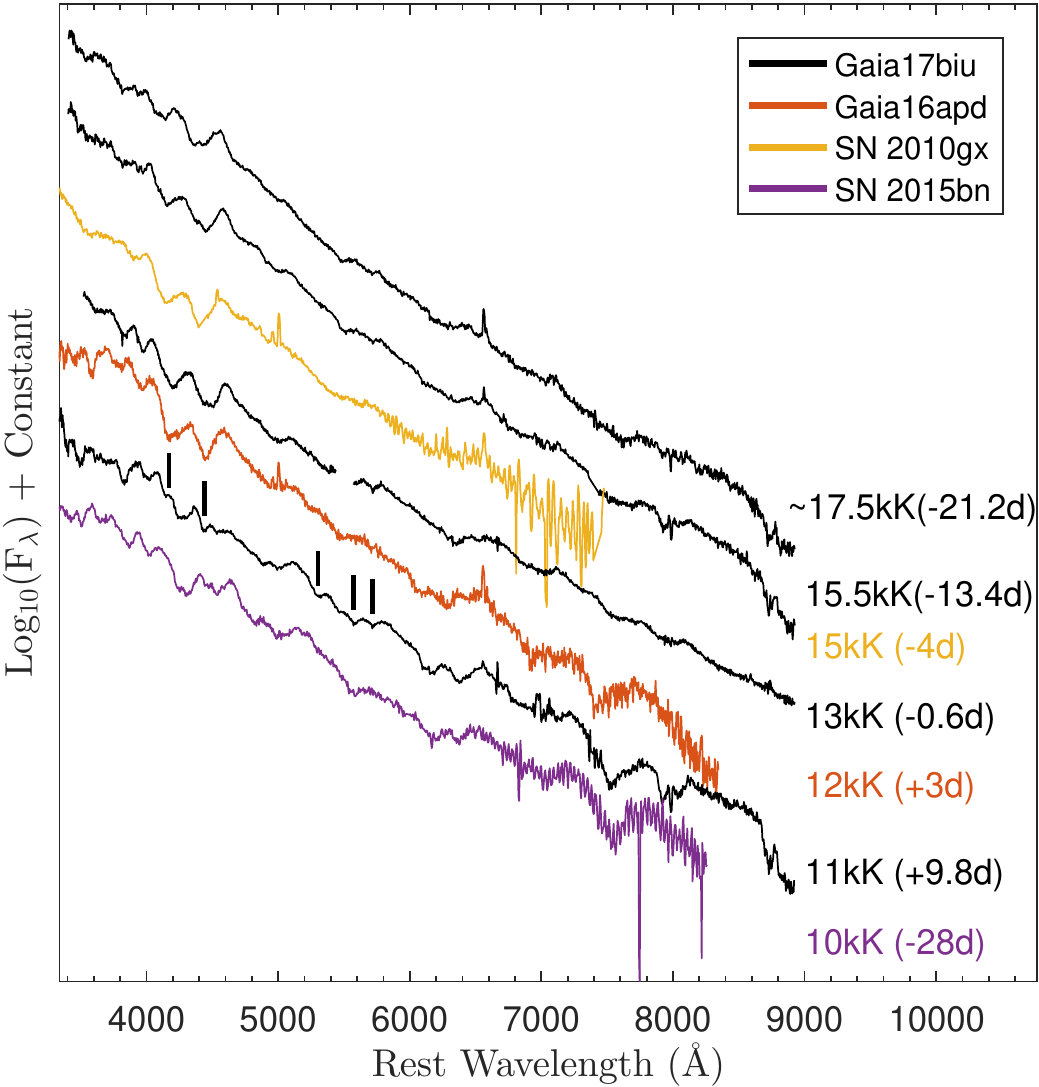}%
	\caption{Rest-frame spectra of \sn\ as compared with spectra of the SLSNe-I SN~2010gx \citep{2010ApJ...724L..16P}, SN 2015bn \citep{2016ApJ...826...39N}, and Gaia16apd 
		\citep{2017MNRAS.469.1246K}. Some weak line 
		features which are not visible in spectra of other SLSNe-I are marked with solid black lines on 
		the day-10 spectrum. In addition to the name of the SN and the epoch of observation,
		each spectrum is labeled by the estimated black-body temperature of the SN in units of $\hbox{kK}=10^3$~K.}
	\label{fig:sp.speccomp}
\end{figure}

In Figure~\ref{fig:sp.speccomp} we compare the spectra of \sn\ to those of other SLSNe-I
at three different phases representing the pre-, near-, and post-peak phases of 
evolution.  We tried to match the \sn\ spectra to other SLSNe-I using a
large number of existing SLSNe-I spectra prepared by \citet{2016arXiv161207321L} as SNID \citep{2007ApJ...666.1024B} templates and available in WiseREP \citep{2012PASP..124..668Y}. In general we found that, spectra corresponding to epochs with comparable
black-body temperatures have the best similarity in spectral features, rather than spectra with comparable phases (relative to maximum light) as is usually done in such comparisons \citep[e.g.,][]{2016arXiv161207321L}. This is illustrated in 
Figure~\ref{fig:sp.speccomp}, where we report the estimated temperature along with
the epoch for each spectrum. 
A good example is that a $\sim 10$~day post-peak 
spectrum of \sn\ is best matched by a $-28$~day pre-peak spectrum 
of SN~2015bn, where both sources have estimated temperatures $\sim 10,000$~K.

SLSNe-I are sometimes divided into fast- and slow-decline populations based
on the post-peak decline rates.  \cite{2016ApJ...826...39N} argue that the 
spectra near peak show differences in several features for the two populations, and \citet{2017MNRAS.469.1246K}
show that Gaia16apd appears to ``bridge'' the two populations in terms of
its decline rate and spectra.  The post-peak decline rate of \sn\ makes
it a member of the slow-decline population, but its pre-peak spectra most
closely resemble those of the prototypical fast-decline SLSN-I SN~2010gx, even while its
post-peak spectra most closely resemble those of the slow decline SLSN-I SN~2015bn.
At peak, \sn\ is spectroscopically similar to Gaia16apd.  This suggests that the
light-curve decline rate is unlikely to be a useful indicator for describing the spectroscopic diversity of the SLSN-I population.

\begin{figure}
	\centering
	\includegraphics[width=\linewidth]{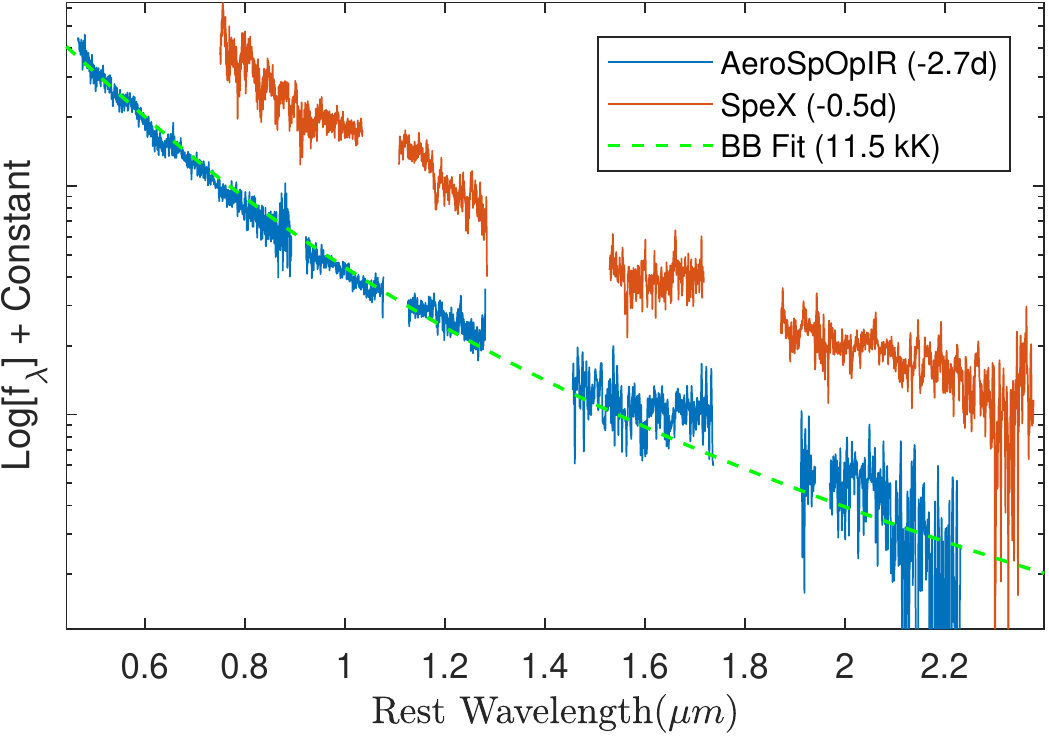}
	\caption{Infrared spectra at two epochs are shown. No prominent features are detectable given the low SNR of these spectra. The continuum follows the Black-Body SED as shown with a dash green line at 11.5 kK temperature. Strong telluric and unreliable regions in the spectra are masked out in the spectra.}\label{fig:IR_spec}
\end{figure}

Figure~\ref{fig:IR_spec} shows the NIR spectra obtained for \sn\ at $ -2.7 $ and $ -0.5 $ days. 
All NIR spectra are smooth, and we do not detect any significant features from either the SN or its host galaxy. \Hei\ \ld10833 is the most prominent feature detected in few SLSNe-I (e.g. Gaia16apd \citealt{2017ApJ...840...57Y} and SN~2012li \citealt{2013ApJ...770..128I}). However, for \sn\ we were unable conclude anything about the presence of \Hei\ due to very strong telluric line contamination in that wavelength range. The SED from the optical through NIR wavelengths shows a continuum consistent with the Rayleigh-Jeans tail of a thermal black body.

\subsection{Spectropolarimetry}  \label{sec:specpol}

The only other SLSN-I with spectropolarimetric observations is SN~2015bn, where \citet{2016ApJ...831...79I} found a significant and increasing degree of polarization between  
$-24$ and $+27$~days. Their results indicated the presence of a consistent dominant axis at both epochs and a strong wavelength dependence of polarization. {Broadband polarimetric observations are available for two SLSNe-I 2015bn \citep{2017ApJ...837L..14L} and LSQ14mo \citep{2015ApJ...815L..10L}. Broadband  polarimetry of SN~2015bn also showed increasing polarization until $ +46 $~day while for LSQ14mo, the broadband polarization during $ -7 $ to $ +19 $~days suggested overall spherical symmetry.}

\begin{figure}
	\centering
	\includegraphics[width=\linewidth]{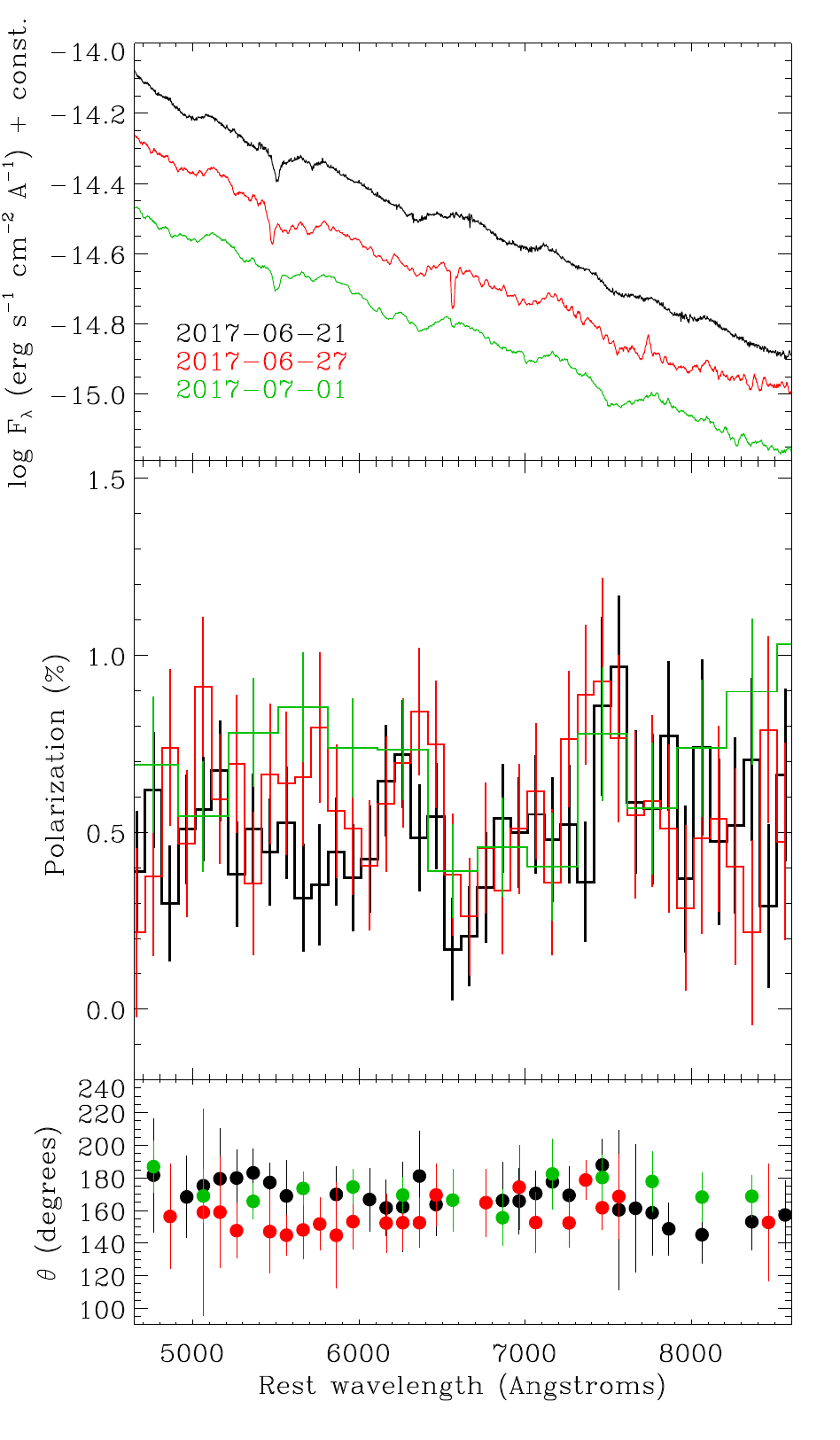}%
	\caption{
		Three epochs of spectropolarimetry for \sn. Top panel: the observed Kast spectra, color coded for each epoch.
		Middle panel: Polarization, given as the rotated Stokes $q$ parameter \citep[see][]{2014MNRAS.442.1166M}. The June and July data have been binned by 100 and 300\,\AA, respectively. Bottom panel: Position angle ($\theta$) for the corresponding epochs, binned to 100\,\AA.
	}
	\label{fig:specpol}
\end{figure}

Our spectropolarimetric results (see Fig.~\ref{fig:specpol}) show that we have detected significant polarization in \sn. To determine whether this polarization is intrinsic to the SN, we must first address the possibility of interstellar polarization (ISP) induced by the dichroic absorption of SN light by interstellar dust grains aligned to the magnetic field of the interstellar medium (ISM). Fortunately, the ISP in the direction of \sn\ appears to be low ($E(B-V) \approx 0.0097\pm0.0005$\,mag; see \S1). According to \cite{1975ApJ...196..261S}, the maximum expected polarization correlates with reddening by $P_{\rm ISP}<9\,E(B-V)$\%, which implies $P_{\rm ISP}<0.09$\% from the Milky Way in the direction of \sn. There are also polarimetric measurements in the literature of a nearby F0~V star, HD\,89536 ($0\fdegr58$ away from \sn), that lies at an estimated spectroscopic-parallax distance of $\sim193$\,pc, sufficiently distant to be useful as a probe of the intervening ISP. The catalogued optical polarization of HD\,89536 is a null detection with $P<0.025$\% \citep{2014A&A...561A..24B}. We thus do not expect significant Galactic ISP in the direction of \sn.

The ISP from the host galaxy is more difficult to ascertain, but the lack of significant \Nai~D absorption features at the rest wavelength of \host\ suggests that the host absorption likely to be less than the low Milky Way value. Furthermore, if the polarization were due to ISP, then we would expect a Serkowski functional form, whereby the polarization peaks near a wavelength of 5500\,\AA\ and drops off at longer wavelengths \citep{1975ApJ...196..261S}. Instead, the average polarization appears to be relatively flat with wavelength, which is more consistent with the effects of electron scattering. We are therefore inclined to interpret the polarized signal as intrinsic to the SN. 

Under the reasonably justified assumption of unsubstantial ISP, it appears that significant intrinsic polarization in the continuum and possible modulations across line features have been detected in \sn. The ``continuum polarization" (integrated over the wavelength range $ 7800-8700 $\,\AA, where the spectra appear to be devoid of line features) is $P_{\rm cont}=0.43\pm0.09$\% at $\theta = 161\pm6^{\circ}$. Taken at face value, the electron-scattering models used by \cite{1991A&A...246..481H} would suggest that this level of polarization is consistent with an ellipsoidal shape on the sky having an axis ratio of $\sim0.9$. Modulations relative to the continuum appear as high as 0.4--0.5\%, particularly in the regions near 6300--6400\,\AA\ and 7300--7600\,\AA. The modulations could thus be associated with blueshifted absorption components of the possible \ion{C}{ii} $\lambda$6580 and O\,{\sc i} $\lambda$7774 lines. The lack of strong deviations in $\theta$ across these features is consistent with global asphericity of the SN atmosphere and its line-forming region, as opposed to a clumpy or irregular line-forming region, which typically results in substantial position-angle changes \citep[see, e.g.,][]{2015MNRAS.453.4467M}. Comparison of the June 21 and 27 data indicate no substantial change in the polarization characteristics between these epochs; slight shifts in polarization and $\theta$ at select wavelengths are near the limit of statistical significance. However, comparison of the June 21 and July 1 data shows a slight indication that the polarization has marginally increased around 5000--6000\,\AA, possibly associated with the Si\,{\sc ii} or C\,{\sc ii} lines. However, the bright night-sky emission lines in this region of the spectrum were particularly strong and rapidly changing, as the observations were performed in substantial twilight, and the polarization increase should be treated with caution.

\section{Radio Observations}

We observed the location of \sn\ at 1.5\,GHz with the electronic Multi-Element Remotely-Linked
Interferometer Network (e-MERLIN) from 2017 June 21 to 23 and with the Karl G. Jansky Very Large Array (VLA)
on 2017 June 22 and 30.  The e-MERLIN observations were made in two continuous runs with a 
bandwidth of 512\,MHz (1254.6--1766.5\,MHz), reduced to $\sim 400$\,MHz after flagging,
using the Knockin, Pickmere, Darnhall, and Cambridge stations along with the Mark~II
(18:00 June 21 to 12:00 June 22) or Defford (15:00 June 22 to 11:00 June 23) stations.
The data were reduced and analysed with the National Radio Astronomy Observatory (NRAO) 
Astronomical Image Processing System ({\sc aips}) following standard procedures. 3C~286 
was used as a flux calibrator and OQ~208 as bandpass calibrator. The phase reference source
J1027$+$4803 (at a projected distance of 2\fdegr11 from the SN position) had a 1.5\,GHz flux density of
147.6\,mJy, which remained constant during the time of the observations. The resolution
was $203 \times 130$\,mas at PA $=-18\fdegr8$. The extended emission of the host galaxy is
resolved out in these observations and we measure a root-mean-square (rms) noise
level of 32\,\mujyb{} at the SN position, corresponding to a 1.5\,GHz
luminosity limit of $< 2\times10^{27}$\,\lunits{} at a $3\sigma$ level.

\begin{figure*}
\includegraphics[width=\linewidth]{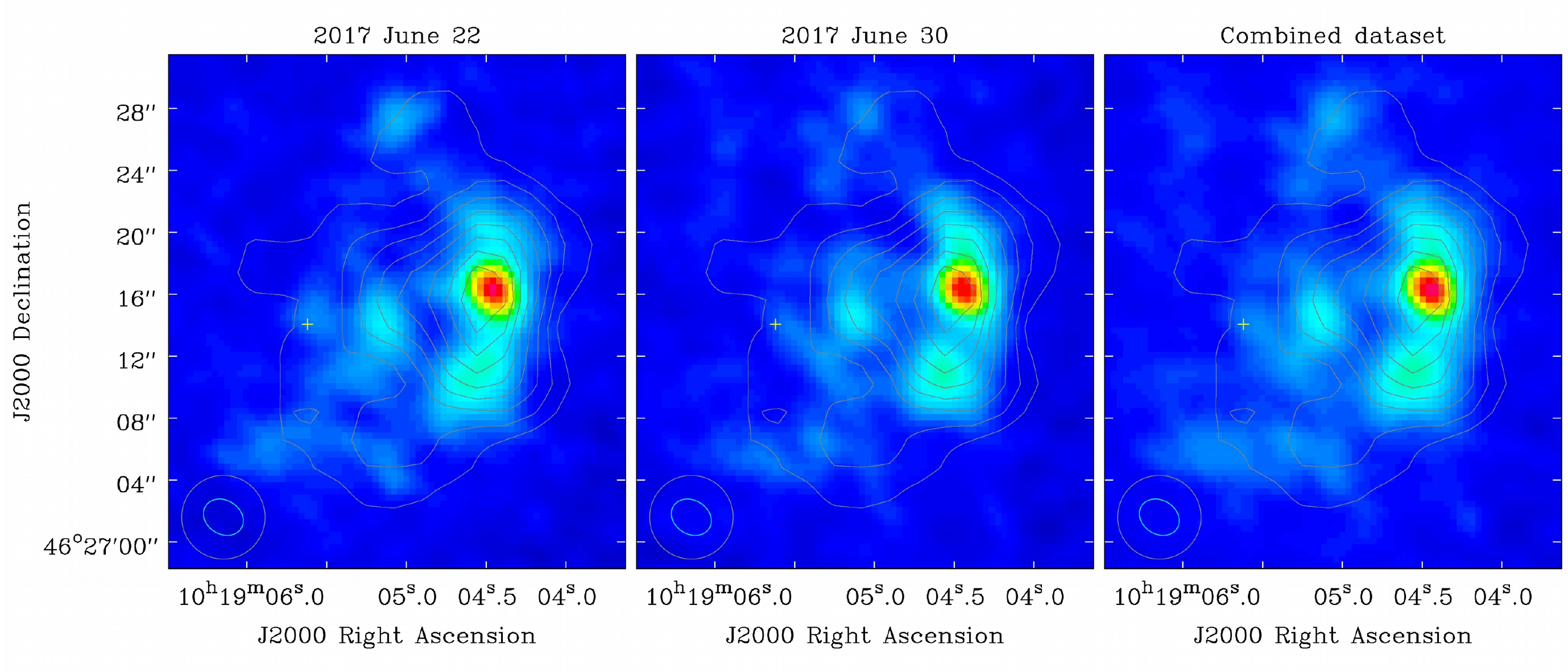}
\caption{NGC\,3191 10\,GHz VLA maps from June 22 (right), June 30 (middle), and from the combined
dataset (right), at a resolution of $2\farcs72\times2\farcs18$, PA $=58\degr{}$. Overlaid in each
map, we show the cutout from the Faint Images of the Radio Sky at Twenty-cm survey
\citep[FIRST;][]{1995ApJ...450..559B} with a full width at half-maximum intensity (FWHM) of $5\farcs4\times5\farcs4$. The beams are shown
in the lower-left corner of each map: the white one corresponds to the recent VLA maps, and
the grey one to the FIRST image cutout. The position of \sn\ is indicated
with a yellow cross.}\label{fig:vla_maps}
\end{figure*}

\sn\ was also observed at a central frequency of 10\,GHz with the VLA on 2017 June 22 and 30.  The data were reduced using
Common Astronomy Software Applications package \citep[{\sc casa};][]{2007ASPC..376..127M} version 4.7.2
with some additional data flagging.  The observations had a total
bandwidth of 4\,GHz with full polarization using 3C~286 as the flux and bandpass calibrator and
J0958$+$4725 (at a projected distance of 3\fdegr67 from \sn) as the phase reference 
source. We achieved rms noise levels of 5.9 and 5.8\,\mujyb{} for the June 22 and 30 epochs,
respectively.  As shown in Figure~\ref{fig:vla_maps} using a common convolving beam 
($2\farcs72\times2\farcs18$, PA $=58\degr{}$) to ease comparisons between the epochs,
the host galaxy is resolved and well detected. 
The brightest structure peaks at $\alpha(\mathrm{J}2000) =  10\hh19\mm04\fs45$ ($\pm 0\farcs03$),
$\delta(\mathrm{J}2000) = 46\degr27\arcmin16\farcs3$  ($\pm 0\farcs03$). This source is coincident
with a strong star-forming region detected in the optical, and with an SDSS spectrum consistent
with an \Hii{} region. The nucleus of the host at
$\alpha(\mathrm{J}2000) =  10\hh19\mm05\fs14$ ($\pm 0\farcs19$),
$\delta(\mathrm{J}2000) = 46\degr27\arcmin14\farcs6$ ($\pm 0\farcs19$) is relatively fainter.
While there is plenty of diffuse emission from the host at the position of the SN, we do not detect any point source at the position of the SN \citep{2017ATel10537....1R} with combined limits
of 23.3\,$\mu$Jy/beam corresponding to $< 5.4 \times 10^{26}$~\lunits\ and no evidence
for variability between the two epochs.

{Radio observations are particularly important as a test for powering SLSNe-I with GRB-like central engines.}
Observations of SN~2015bn 238 days after maximum light placed an upper limit of $< 2\times10^{28}$\,\lunits{}, ruling out its association with a typical long GRB and various off-axis geometries \citep{2016ApJ...826...39N}. However, owing to the late phase of the radio observations, they were not able to place meaningful constraints on an association with the low-luminosity GRBs (possibly not highly collimated) that dominate the local GRB rate. 

In the case of \sn, we have the advantage of proximity (factor of 3.7 closer than SN 2015bn), allowing us to put a tighter upper limit on the radio luminosity of $< 5.4\times10^{26}$\,\lunits{} at 10\,GHz. Furthermore, our radio observations were taken close to the optical peak, when the radio afterglow luminosity would also be expected to be near maximum for low-luminosity GRBs. Using the radio upper limit of \sn, we can rule out its association with low-luminosity GRBs across most of the observed luminosities. When low-luminosity GRBs have been observed in the radio \citep[e.g.,][]{1998Natur.395..670G, 2004Natur.430..648S, 2006Natur.442.1014S, 2013ApJ...778...18M}, their 8.5\,GHz luminosities are generally $10^{28}$--$10^{29}$\,\lunits{} at phases similar to those when \sn\ was at $<5.4\times10^{26}$\,\lunits{}, with the exception of GRB 060218 at $\sim 10^{27}$\,\lunits{}. This appears to largely rule out an association of this SLSN-I with a GRB radio afterglow.

Comparing our radio upper limit around day 30 after explosion with typical radio fluxes
of SNe Ibc \citep{2010Natur.463..513S}, we find that \sn\ must have been a weaker source than most SNe Ibc.
A normal SNe Ibc with a spectrum that peaks at 10 GHz around 30 days has a flux 
of $ \sim 2\times10^{27} $\,\lunits{}, significantly higher than our upper limit of $ 5.4\times10^{26} $\,\lunits{}. However, judging from 
\cite{2010Natur.463..513S}, if the spectral peak at this epoch was below  $ \sim 3 $ GHz, an SN Ibc would most likely go
undetected in our data. Likewise, for a spectral peak above  $ \sim 20 $ GHz at 30 days, synchrotron self-absorption 
would make a detection unlikely. Thus, our 10 GHz data cannot rule out that \sn\ could be an SN Ibc-like radio source, although it would be among the weakest in this class. For example, SNe 2003gk 
\citep{2014MNRAS.440..821B} and 2014C \citep{2017ApJ...835..140M} would both have been undetected at 30 days given our upper
limits, despite these supernovae being much brighter at later epochs.

{The radio observations presented in this paper place stringent upper limits on the radio emission from this source, showing no evidence for strong interactions of the ejecta with the CSM at this point in its evolution. Further deep radio observations are required to determine if interaction between the eject and the CSM at later times may result in greater levels of radio emission due to relatively dense CSM.}

\section{X-ray Observations}

\textit{Swift} also observed the field of \sn\ with its X-ray telescope \citep[XRT; ][]{2005SSRv..120..165B} for a total of 33,661\,s. All observations were performed in photon counting mode \citep[PC mode;][]{2004SPIE.5165..217H} and were processed in the standard way by running {\it xrtpipeline}. The resulting event files were then combined in {\it XSELECT} in order to obtain spectra, event files, and images. X-ray positions were determined by using the online XRT product tool at the University of Leicester website\footnote{http://www.swift.ac.uk/user\_objects}.  For the average X-ray spectrum we created an auxilliary response file (arf) for each single observation using the task {\it xrtmkarf} and 
combined them into a single arf by using the FTOOL task {\it addarf}. We used the XRT pc mode response file {\it swxpc0to12s6\_20130101v014.rmf}. The spectral analysis was performed using  XSPEC version 12.8.2 \citep{1985MNRAS.217..105A}.

After coadding the data for the first two weeks of \textit{Swift} observations, we noticed enhanced X-ray emission close to the optical position of \sn. This period had a total exposure time of 13\,ks \citep{2017ATel.10499...1G}. We measured the position of this X-ray source to be $\alpha_{J2000} = 10^{\rm h}19^{\rm m}05.\!\!^{\rm s}77$ and $\delta_{J2000} = +46^{\circ}27'14\farcs1$ with an uncertainty of $5\farcs1$. This position was $4\farcs6$ away from the optical counterpart of \sn\ and $7\farcs3$ from the center of NGC~3191, the host galaxy of \sn. Applying the Bayesian method described by \citet{1991ApJ...374..344K}, we obtained a count rate in the 0.3--10\,keV energy range of $(9.5^{+3.2}_{-2.7}\times 10^{-4}$\,counts\,s$^{-1})$, which corresponds to a flux in the 0.3--10\,keV band of $4.1^{+1.6}_{-1.4}\times 10^{-14}$\,erg\,s$^{-1}$\,cm$^{-2}$. Assuming that this X-ray source is located at the distance of NGC 3191 ($D_L = 139$\, Mpc), this corresponds to a luminosity of $10^{41}$\,erg\,s$^{-1}$. 

Although the X-ray position obtained over the first two weeks of \textit{Swift} observations initially suggested that this X-ray source may well be the counterpart of \sn, adding more observations in the following weeks made this conclusion less convincing \citep{2017ATel.10531...1G}. Our new analysis included all available data obtained between 2017 June 2 and July 4. The X-ray spectrum of the X-ray source can be fitted by a single power-law model with the absorption column density fixed to the Galactic value \citep[$N_{\rm H}=9.39\times 10^{19}$\,cm$^{-2}$][]{2005A&A...440..775K}, a photon index $\Gamma = 1.88^{+0.51}_{-0.49}$, and a flux in the observed 0.3--10\,keV band of 
$2.9^{+1.2}_{-0.6}\times 10^{-14}$\,erg\,s$^{-1}$\,cm$^{-2}$. 
The count rate obtained from these data is $9.0^{+1.9}_{-1.7}\times 10^{-4}$\,counts\,s$^{-1}$. There is no evidence for any variability of the X-ray source. The source position in this 34\,ks observation is $\alpha_{J2000} = 10^{\rm h}19^{\rm m}04.\!\!^{\rm s}96$ and $\delta_{J2000} = +46^{\circ}27'15\farcs8$ with an uncertainty of $6\farcs4$. This position is $7\farcs0$ away from the optical position of \sn\ and $1\farcs7$ from the position of NGC 3191. This new position suggests that the X-ray emission is likely associated with the starburst region in the center of NGC 3191 and not coming from \sn.

{We also obtained a 3$\sigma$ upper limit at the optical position of the supernova in the 0.3-10 keV range. We extracted source counts in a circular region with a radius of 2 pixels (equivalent to $ 4\farcs7 $) centered on the optical position of \sn. The background was subtracted from an annulus with an inner radius of 3$^{''}$ and an outer radius of 10$^{''}$. Without PSF correction, we obtained a 3$\sigma$ upper limit of 3.4$\times 10^{-3}$ counts s$^{-1}$ applying the Bayesian method by \citet{1991ApJ...374..344K}. Assuming the same spectral model as above, this corresponds to a flux limit in the 0.3-10 keV band of 1.1$\times 10^{-13}$ ergs s$^{-1}$ cm$^{-2}$ which assuming the luminosity distance of NGC 3191 converts to a luminosity limit of 2.5$\times 10^{41}$ ergs s$^{-1}$.  Due to the broad PSF of the {\it Swift} XRT, we caution this limit may be weakened by strong contamination from the nearby star forming region. }
A secure X-ray constraint at the 0.3-10 keV band around bolometric maximum could potentially test whether there are circumstellar interactions \citep[e.g.,][]{2012MNRAS.419.1515D}.

\section{Summary and Discussion} \label{sec:discuss}

In summary, we identify \sn\ as the the lowest redshift SLSN-I to date, exploding in a massive and metal-rich host galaxy that is typical of ccSNe but atypical of most known SLSNe-I. Previously, it was suggested that SLSN-I production might be strongly suppressed at high metallicities \citep[e.g.,][]{2016arXiv161205978S}, and the purported requirement for a metal-poor environment was seen as evidence supporting the birth of a fast-spinning magnetar as the central engine for SLSNe-I \citep[e.g.,][]{2016ApJ...830...13P}. But the relatively high volume rate implied by the close distance of \sn\ suggests that any metallicity effect on SLSNe-I production rate is weaker than presently believed.

The curious fact that the two all-sky surveys for bright transients, ASAS-SN 
and Gaia, have both found SLSNe-I in massive, higher metallicity galaxies 
demands explanation.  It is difficult to explain as a selection 
effect in ASAS-SN or Gaia, since almost every observational selection effect 
in an untargeted transient survey favors higher survey efficiencies in 
less luminous galaxies.  A selection effect against non-dwarf galaxies 
in higher redshift surveys seems more likely.  For example, the discovery 
rate of tidal disruption events relative to Type~Ia SNe 
\citep{2016MNRAS.455.2918H} and the radial distribution of SNe relative to the
centers of galaxies in ASAS-SN \citep{2017MNRAS.467.1098H} clearly show that 
both amateurs and most professional surveys have been strongly biased 
against identifying transients close ($\sim$kpc) to the central regions of 
luminous galaxies where both ASASSN-15lh and Gaia17biu were discovered. 
For example, the median offset of Type~Ia SNe 
in PTF is about $5$\,kpc \citep{2017ApJ...836...60L}, as compared to a 
median of $2.6$\,kpc in ASAS-SN \citep{2017arXiv170402320H}, and the 
$ 3 $~kpc offset of \sn.  This incompleteness is likely a combination 
of the additional systematic problems in detecting transients in the central 
regions of luminous galaxies and a human bias against making expensive 
spectroscopic observations of candidate transients
in regions with high false positive rates.

\begin{figure}
	\centering
	\includegraphics[width=0.8\linewidth]{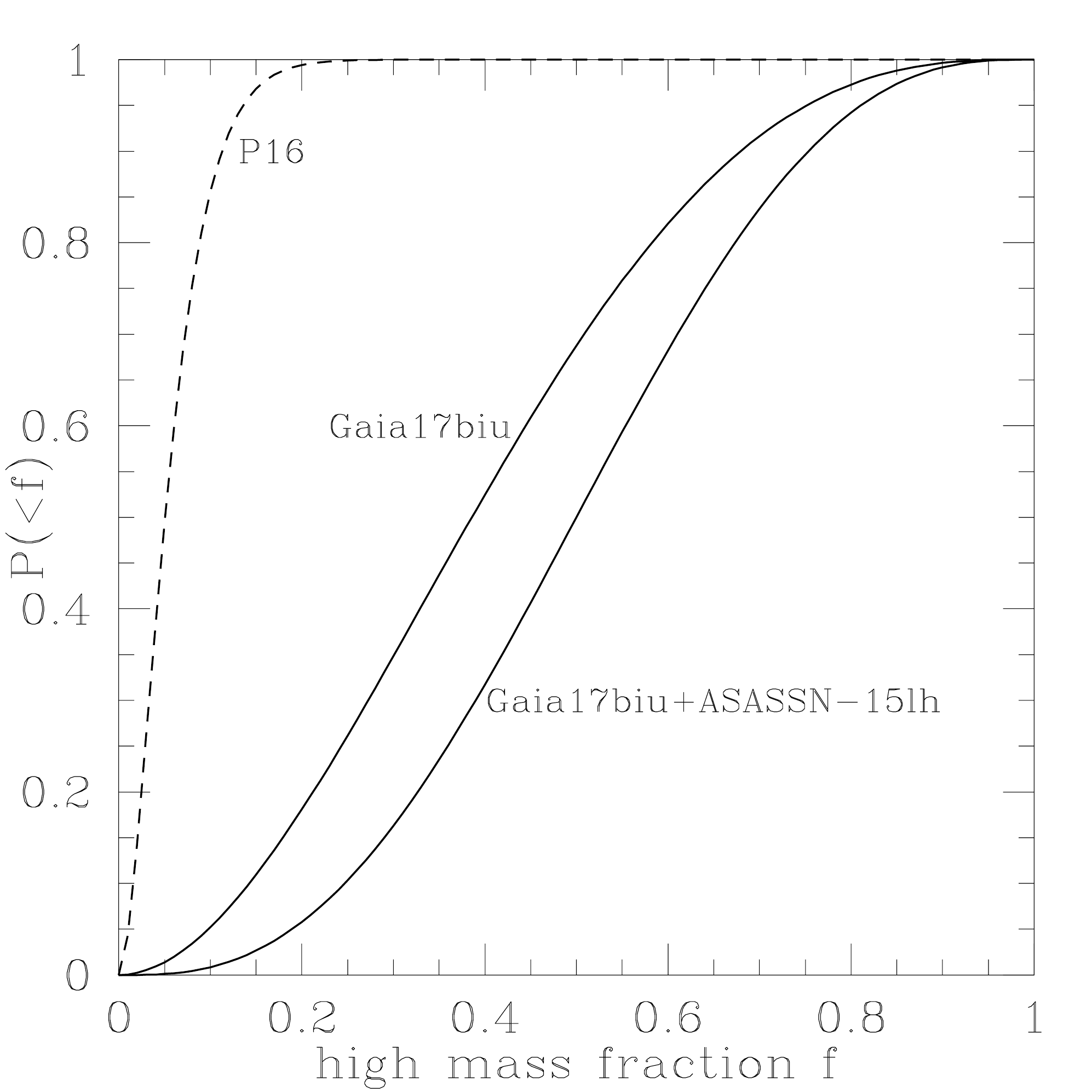}
	\caption{The integral probability distribution for fraction of high mass hosts of SLSNe-I. The dashed line shows the probability distribution of the P16 sample while the solid lines show the probability distribution for Gaia/ASAS-SN sample, with and without including ASASN-15lh.}\label{fig:rate_prob}
\end{figure}

We can roughly quantify the problem by assuming that the host stellar mass 
distribution from P16 is representative and that the surveys differ only 
in their effective survey volumes due to the differences in photometric 
depth.  Under these assumptions, the SLSNe-I host mass and metallicity 
distributions in Gaia (or ASAS-SN) should be the same as in PTF (P16).  
Gaia has found two
SLSN-I, \sn\ and Gaia16apd \citep{2017MNRAS.469.1246K}, where the latter is in a low mass host.
The ASAS-SN sample also includes two, SN~2015bn, which was discovered by 
PS1 \citep{2015ATel.7153....1H} but recovered by ASAS-SN, and ASASSN-15lh \citep{2016Sci...351..257D}.  SN~2015bn
is in a low mass host, while ASASSN-15lh is in a high mass host.  
We carry out our calculations both with and without ASASSN-15lh
since its identification as an SLSN-I is debated 
\citep{2016Sci...351..257D, 2016NatAs...1E...2L, 2017MNRAS.466.1428G}. 
	
{The P16 sample contains 32 SLSN-I, one of which is in a high mass
host.  The probability for the fraction $f$ in high mass hosts ($ M_*\gtrsim10^{10} \msun$) is 
simply the binomial distribution $P(f) \propto f (1-f)^{31}$, and
we show the resulting integral probability distribution for $f$ 
in Figure.~\ref{fig:rate_prob}.  The median estimate is $f=0.050$ with a 90\%
confidence region of $0.011 < f < 0.14$. The Gaia and ASAS-SN
low redshift surveys have either $P(f) \propto f (1-f)^2$ without
ASASSN-15lh or $f^2 (1-f)^2$ if it is included.  The integral
distributions for these two cases are also shown in Fig.~\ref{fig:rate_prob},
and we see there is very little overlap. 
The medians for the low redshift samples are $f=0.39$ ($0.093 < f < 0.75$) 
without ASASSN-15lh and $f=0.50$ ($0.19 < f < 0.81$) with ASASSN-15lh.  
Alternatively, we can average the probabilities of finding one (two)
or more SLSN-I in high mass galaxies in the Gaia/ASAS-SN samples
over the probability of $f$ implied by the P16 sample, to find that 
there is only a 16\% (2.5\%) probability of such a result.  These
are not low enough likelihoods to be definitive, but combined with
the evidence that the higher redshift surveys are biased against
events as close to galactic centers as \sn\ and ASASSN-15lh, there
is certainly a strong suggestion that the prevalence of SLSN-I in
high mass galaxies is being underestimated.  If we ignore the 
question of incompleteness, simply combining the low redshift
samples with P16 raises the median to $f=0.074$ ($0.098$)
with a 90\% confidence range of $0.023<f< 0.16$ ($0.038 < f < 0.20$)
excluding (including) ASASSN-15lh.}

The proximity and high apparent brightness of \sn\ allowed us to carry out intensive and detailed follow-up observations during its early phases across a wide range of wavelengths. Its peak luminosity of $M_g \simeq -21$\,mag is {typical of known SLSN-I luminosity distribution \citep{2017arXiv170801623D,2017arXiv170801619L}}, and we find that its fast rise time is consistent with an empirical correlation between optical luminosity and rise time for well-observed SLSNe-I. We see no evidence for an early ``bump'' or undulations as seen in several other SLSNe-I. Both its rise to the peak and decline from the peak follow strikingly simple exponential forms, with a rapid reversal at the peak, and these features may help to distinguish theoretical models of powering mechanisms \citep[e.g.,][]{2013ApJ...773...76C}.

While \sn\ is a slowly declining SLSN-I, its spectroscopic resemblance to both the fast- and slow-declining sub-classes at different phases suggests that such a division may not be a useful description of the spectroscopic diversity of SLSNe-I. Our unprecedented high-SNR spectra also reveal several new features in SLSNe-I, potentially shedding new light into the chemical composition of the ejecta.
{We also identified new and subtle spectral features which are short-lived and these detections were possible due to high cadence and high SNR spectroscopic observations.
Apart from the extraordinarily linear rising and declining light curve, the photometric and spectroscopic features are broadly similar to other well-observed SLSNe-I. Given the very limited number of well-observed SLSNe, the uniquely identified features in \sn\ may not be characterized as unusual, rather this possibly adds to the diversity of SLSNe-I.}

In addition, we obtained spectropolarimetric observations, rare for a SLSN-I, showing that the 
ejecta are consistent with a global departure from spherical symmetry, the true extent of which is dependent on the uncertain viewing angle.
Our uniquely tight constraint on its radio luminosity largely rules out an association of \sn\ with the GRB mechanism across the known luminosity function of radio afterglows.

The explosion model and energy supply mechanism for SLSNe-I are not known. Some proposed models include the spindown of highly magnetized, fast-rotating neutron stars \citep{1974ApJ...191..465B, 2010ApJ...717..245K, 2010ApJ...719L.204W} or quark stars \citep{2016ApJ...817..132D}, pair-instability explosions \citep{1967PhRvL..18..379B}, and ejecta interactions with circumstellar material (CSM) \citep[e.g.][]{2010arXiv1009.4353B,2016ApJ...829...17S}. Thanks to its proximity, \sn\ is likely observable to very late evolution phases. Combined with the early-time data presented here, such late-time observations, especially the nebular-phase spectra, can help test theoretical models and clarify the chemical composition and ejecta structure. Also late time radio observations of \sn\ will be able to place strong constraints on possible associations with off-axis GRBs \citep[e.g.,][]{2002ApJ...576..923L}.

\acknowledgements

We thank David Sand, Carl Melis, Mike Celkins, Richard Puetter, K. Crawford and M. Varakianfor their help with some of the observations and the Lick Observatory staff for their assistance. We are grateful to Boaz Katz, Kuntal Misra, A. Gal-Yam and S. Schulze for helpful comments.

S.B., S.D., and P.C. acknowledge Project 11573003 supported by NSFC. S.B. is partially supported by China postdoctoral science foundation grant No. 2016M600848. A.P., L.T., S.B., and N.E.R. are partially supported by the PRIN-INAF 2014 project Transient Universe: unveiling new types of stellar explosions with PESSTO.
C.S.K., K.Z.S., and T.A.T. are supported by US National Science Foundation (NSF) grants AST-1515927 and AST-1515876. 
A.V.F.'s supernova research group at U.C. Berkeley is grateful for                     
generous financial assistance from the Christopher R. Redlich Fund,                      
the TABASGO Foundation, Gary and Cynthia Bengier (T.d.J. is a Bengier                    
Postdoctoral Fellow), and the Miller Institute for Basic Research                         
in Science (U.C. Berkeley).
We acknowledge support by the Ministry of Economy, Development, and Tourism's Millennium Science Initiative through grant IC120009, awarded to The Millennium Institute of Astrophysics, MAS, Chile (J.L.P., C.R.-C.) and from CONICYT through FONDECYT grants 3150238 (C.R.-C.) and 1151445 (J.L.P.).
B.J.S. is supported by NASA through Hubble Fellowship grant HST-HF-51348.001 awarded by the Space Telescope Science Institute, which is operated by the Association of Universities for Research in Astronomy, Inc. under contract NAS 5-26555. T.W.-S.H is supported by the DOE Computational Science Graduate Fellowship, grant number DE-FG02-97ER25308. E.Y.H., C.A., S.D., and M.S. acknowledge support provided by NSF grants AST-1008343 and AST-1613472 and by the Florida Space Grant Consortium.
B.S.G. and DEMONEXT were partially funded by NSF CAREER grant AST-1056524.
G.H. is supported by NSF grant AST-1313484.
M.D.S. is supported by a research grant (13261) from VILLUM FONDEN. NUTS is supported in part by the Instrument Centre for Danish Astrophysics (IDA).
Work by S.V.Jr. is supported by NSF Graduate Research Fellowship under grant DGE-1343012. The work of N.E.R. was completed at the {\it Institut de Ci\`encies de l'Espai} at the Autonomous University of Barcelona's Campus; she thanks the Institute for its hospitality. S.H. is supported by the Young Scholars Program of Shandong University, Weihai.
PJB's work on SLSNe is supported by the Swift GI program through grant NNX15AR41G. J.H.W. acknowledges the support by the National Research Foundation of Korea grant funded by the Korea government (No.2017R1A5A1070354). MF is supported by a Royal Society - Science Foundation Ireland University Research Fellowship. C.G. is supported by the Carlsberg Foundation.

This research was made possible through the use of the AAVSO Photometric All-Sky Survey (APASS), funded by the Robert Martin Ayers Sciences Fund.
We acknowledge ESA Gaia, DPAC, and the Photometric Science Alerts Team (http://gsaweb.ast.cam.ac.uk/alerts).
We thank {\it Swift} Acting PI S. Bradley Cenko, the Observation Duty Scientists, and the science planners for promptly approving and executing our observations. 
This research uses data obtained through the Telescope Access Program (TAP), which has been funded by ``the Strategic Priority Research Program-The Emergence of Cosmological Structures" of the Chinese Academy of Sciences (Grant No.11 XDB09000000) and the Special Fund for Astronomy from the Ministry of Finance.
This work made use of the data products generated by the NYU SN group, and released under DOI:10.5281/zenodo.58766,
available at \url{https://github.com/nyusngroup/SESNtemple/} 

Research at Lick Observatory is partially supported by a very generous 
gift from Google, as well as by contributions from numerous individuals 
including                                     
Eliza Brown and Hal Candee,                                                 
Kathy Burck and Gilbert Montoya,                                            
David and Linda Cornfield,                                                  
William and Phyllis Draper,                                                 
Luke Ellis and Laura Sawczuk,                                             
Alan and Gladys Hoefer,                                                      
Roger and Jody Lawler,
DuBose and Nancy Montgomery,                                               
Jeanne and Sanford Robertson,                                              
Stanley and Miriam Schiffman,                                              
Thomas and Alison Schneider,                                               
the Hugh Stuart Center Charitable Trust,                               
Mary-Lou Smulders and Nicholas Hodson,                                   
Clark and Sharon Winslow,                                                  
Weldon and Ruth Wood,                                                 
and many others.
A major upgrade of the Kast spectrograph on the Shane 3~m telescope
at Lick Observatory was made possible through generous gifts from
William and Marina Kast as well as the Heising-Simons Foundation.

Observations made with the Nordic Optical Telescope, operated by the Nordic Optical Telescope Scientific Association, and with the Gran Telescopio Canarias (GTC), at the Observatorio del Roque de los Muchachos, La Palma, Spain, of the Instituto de Astrofisica de Canarias. The data presented here were obtained in part with ALFOSC, which is provided by the Instituto de Astrofisica de Andalucia (IAA) under a joint agreement with the University of Copenhagen and NOTSA.

We thank the Las Cumbres Observatory and its staff for its continuing support of the ASAS-SN project. We are grateful to M. Hardesty of the OSU ASC technology group. ASAS-SN is supported by the Gordon and Betty Moore Foundation through grant GBMF5490 to the Ohio State University and NSF grant AST-1515927. Development of ASAS-SN has  been supported by NSF grant AST-0908816, the Mt. Cuba Astronomical Foundation, the Center for Cosmology and AstroParticle Physics at the Ohio State University, the Chinese Academy of Sciences South America Center for Astronomy (CAS- SACA), the Villum Foundation, and George Skestos. This paper uses data products produced by the OIR Telescope Data Center, supported by the Smithsonian Astrophysical Observatory. 

The radio observations were carried out as e-MERLIN and VLA (17A-464/AR961, 17A-470/AR962) director's discretionary observations. e-MERLIN is a National Facility operated by the University of Manchester at Jodrell Bank Observatory on behalf of STFC. The National Radio Astronomy Observatory is a facility of the NSF operated under cooperative agreement by Associated Universities, Inc.

The Liverpool Telescope is operated on the island of La Palma by Liverpool John Moores University in the Spanish Observatorio del Roque de los Muchachos of the Instituto de Astrofisica de Canarias with financial support from the UK Science and Technology Facilities Council (STFC).

Partially based on observations collected at Copernico telescope (Asiago, Italy) of the INAF -- Osservatorio Astronomico di Padova, and on observations collected with the 1.22 m Galileo telescope of the Asiago Astrophysical Observatory, operated by the Department of Physics and Astronomy ``G. Galilei'' of the Universit\`a of Padova.
 
 This work was supported in part by The Aerospace Corporation's Technical Investment Program.

\software{MATLAB, Python, IDL, \synow\ \citep{1997ApJ...481L..89F,1999MNRAS.304...67F,2002ApJ...566.1005B}, Astropy \citep{2013A&A...558A..33A},  CASA \citep[v4.72;][]{2007ASPC..376..127M}, HEASOFT \citep{2014ascl.soft08004N}, IRAF \citep{1993ASPC...52..173T}, LT pipeline \citep{2012AN....333..101B,2014SPIE.9147E..8HP}, DoPHOT \citep{1993PASP..105.1342S}, \textsc{FOSCGUI}}

\bibliography{ms}

\input{./photsn.tex}
\input{./TR.tex}
\input{./speclog.tex}
\end{document}

%% file: photsn.tex

\begin{center}
	

      \tablecaption{Photometry of \sn.\label{tab:photsn_simple}}

      \tabletypesize{\fontsize{2.8mm}{3.0mm}\selectfont}

{ \startlongtable
  \begin{deluxetable*}
  	{c c r c c c c c l}	
  	\tablecaption{Photometry of \sn.\label{tab:photsn_simple}}
  	
  	\tabletypesize{\fontsize{2.8mm}{3.0mm}\selectfont}
  	
\tablehead{
            UT Date &         JD $-$ &  Phase$^a$ &             $B$ &                  $g$ &                   $V$ &                  $r$ &                  $i$ &  Telescope$ ^b $ \\
      &  2,457,000 & (days) &              (mag) &              (mag) &               (mag) &              (mag) &              (mag) &        / Inst.     }

          \startdata              
2017-05-16.29 & 889.79 & $-35.43$ &                --- &                --- & $ > $19.090         &                --- &                --- &         AS \\
2017-05-18.31 & 891.81 & $-33.47$ &                --- &                --- &  $ > $18.760        &                --- &                --- &         AS \\
2017-05-20.26 & 893.76 & $-31.58$ &                --- &                --- &  17.360 $\pm$ 0.140 &                --- &                --- &         AS \\
2017-05-21.30 & 894.80 & $-30.56$ &                --- &                --- &  18.710 $\pm$ 0.500 &                --- &                --- &         AS \\
2017-05-22.26 & 895.76 & $-29.63$ &                --- &                --- &  16.970 $\pm$ 0.110 &                --- &                --- &         AS \\
2017-05-24.33 & 897.83 & $-27.63$ &                --- &                --- &  16.800 $\pm$ 0.120 &                --- &                --- &         AS \\
2017-05-25.25 & 898.75 & $-26.73$ &                --- &                --- &  16.860 $\pm$ 0.100 &                --- &                --- &         AS \\
2017-05-26.25 & 899.75 & $-25.76$ &                --- &                --- &  16.210 $\pm$ 0.070 &                --- &                --- &         AS \\
2017-05-27.28 & 900.78 & $-24.77$ &                --- &                --- &  16.380 $\pm$ 0.080 &                --- &                --- &         AS \\
2017-05-28.16 & 901.66 & $-23.91$ &                --- & 16.123 $\pm$ 0.057 &                 --- & 16.354 $\pm$ 0.051 & 16.529 $\pm$ 0.069 &         Io \\
2017-05-29.16 & 902.66 & $-22.94$ & 16.060 $\pm$ 0.034 &                --- &  16.136 $\pm$ 0.043 & 16.296 $\pm$ 0.035 & 16.461 $\pm$ 0.055 &        LCO \\
2017-05-29.18 & 902.68 & $-22.92$ &                --- & 16.014 $\pm$ 0.058 &                 --- & 16.187 $\pm$ 0.057 &                --- &         Io \\
2017-05-30.16 & 903.66 & $-21.97$ &                --- & 16.057 $\pm$ 0.153 &                 --- &                --- &                --- &         Io \\
2017-05-31.90 & 905.40 & $-20.29$ & 15.895 $\pm$ 0.096 & 15.796 $\pm$ 0.028 &  15.899 $\pm$ 0.051 & 16.103 $\pm$ 0.023 & 16.350 $\pm$ 0.041 &         LT \\
2017-06-01.18 & 905.68 & $-20.01$ & 16.008 $\pm$ 0.159 & 15.881 $\pm$ 0.102 &  15.849 $\pm$ 0.031 & 16.063 $\pm$ 0.065 & 16.232 $\pm$ 0.086 & DN,Io,PO \\
2017-06-02.20 & 906.70 & $-19.02$ & 15.800 $\pm$ 0.029 & 15.745 $\pm$ 0.043 &  15.780 $\pm$ 0.030 & 15.958 $\pm$ 0.041 & 16.119 $\pm$ 0.046 &       PO \\
2017-06-02.25 & 906.75 & $-18.97$ &                --- & 15.797 $\pm$ 0.062 &                 --- & 15.983 $\pm$ 0.059 & 16.120 $\pm$ 0.076 &         Io \\
2017-06-03.20 & 907.70 & $-18.05$ & 15.806 $\pm$ 0.031 & 15.748 $\pm$ 0.024 &  15.775 $\pm$ 0.023 & 15.947 $\pm$ 0.037 & 16.067 $\pm$ 0.047 &       PO \\
2017-06-04.20 & 908.70 & $-17.08$ & 15.766 $\pm$ 0.031 & 15.685 $\pm$ 0.045 &  15.724 $\pm$ 0.025 & 15.879 $\pm$ 0.045 & 15.944 $\pm$ 0.087 &       PO \\
2017-06-04.31 & 908.81 & $-16.98$ &                --- & 15.645 $\pm$ 0.076 &                 --- &                --- &                --- &         Io \\
2017-06-05.20 & 909.70 & $-16.11$ & 15.673 $\pm$ 0.070 &                --- &  15.592 $\pm$ 0.061 & 15.826 $\pm$ 0.091 & 15.996 $\pm$ 0.106 &         DN \\
2017-06-05.21 & 909.71 & $-16.10$ & 15.715 $\pm$ 0.027 & 15.626 $\pm$ 0.036 &  15.644 $\pm$ 0.029 & 15.818 $\pm$ 0.038 & 15.969 $\pm$ 0.044 &       PO \\
2017-06-05.94 & 910.44 & $-15.39$ &                --- & 15.510 $\pm$ 0.030 &  15.569 $\pm$ 0.052 & 15.731 $\pm$ 0.024 & 15.975 $\pm$ 0.037 &         LT \\
2017-06-06.20 & 910.70 & $-15.14$ & 15.610 $\pm$ 0.070 &                --- &  15.582 $\pm$ 0.059 & 15.821 $\pm$ 0.090 & 15.926 $\pm$ 0.115 &         DN \\
2017-06-06.22 & 910.72 & $-15.12$ & 15.652 $\pm$ 0.029 & 15.573 $\pm$ 0.032 &  15.565 $\pm$ 0.031 & 15.756 $\pm$ 0.049 & 15.884 $\pm$ 0.050 &       PO \\
2017-06-07.19 & 911.69 & $-14.18$ & 15.549 $\pm$ 0.078 &                --- &  15.447 $\pm$ 0.071 & 15.685 $\pm$ 0.093 & 15.940 $\pm$ 0.123 &         DN \\
2017-06-07.22 & 911.72 & $-14.15$ & 15.587 $\pm$ 0.033 & 15.483 $\pm$ 0.038 &  15.495 $\pm$ 0.027 & 15.666 $\pm$ 0.029 & 15.816 $\pm$ 0.048 &       PO \\
2017-06-08.96 & 913.46 & $-12.47$ & 15.471 $\pm$ 0.142 & 15.314 $\pm$ 0.052 &  15.359 $\pm$ 0.053 &                --- &                --- &         LT \\
2017-06-09.13 & 913.63 & $-12.30$ & 15.347 $\pm$ 0.094 &                --- &  15.413 $\pm$ 0.078 & 15.576 $\pm$ 0.078 &                --- &        LCO \\
2017-06-09.20 & 913.70 & $-12.23$ & 15.409 $\pm$ 0.072 &                --- &  15.354 $\pm$ 0.077 & 15.515 $\pm$ 0.089 & 15.762 $\pm$ 0.115 &         DN \\
2017-06-10.17 & 914.67 & $-11.29$ &                --- & 15.274 $\pm$ 0.069 &                 --- & 15.450 $\pm$ 0.062 & 15.616 $\pm$ 0.086 &         Io \\
2017-06-10.18 & 914.68 & $-11.28$ & 15.366 $\pm$ 0.079 &                --- &  15.265 $\pm$ 0.058 & 15.450 $\pm$ 0.083 & 15.708 $\pm$ 0.088 &         DN \\
2017-06-10.20 & 914.70 & $-11.26$ & 15.326 $\pm$ 0.026 & 15.262 $\pm$ 0.043 &  15.259 $\pm$ 0.035 & 15.430 $\pm$ 0.055 & 15.586 $\pm$ 0.059 &       PO \\
2017-06-11.17 & 915.67 & $-10.32$ &                --- &                --- &  15.202 $\pm$ 0.042 & 15.394 $\pm$ 0.038 & 15.528 $\pm$ 0.056 &        LCO \\
2017-06-11.18 & 915.68 & $-10.31$ & 15.238 $\pm$ 0.068 &                --- &  15.158 $\pm$ 0.056 & 15.361 $\pm$ 0.081 & 15.610 $\pm$ 0.081 &         DN \\
2017-06-11.20 & 915.70 & $-10.29$ &                --- & 15.176 $\pm$ 0.063 &                 --- & 15.361 $\pm$ 0.053 & 15.543 $\pm$ 0.066 &         Io \\
2017-06-11.20 & 915.70 & $-10.29$ & 15.262 $\pm$ 0.026 & 15.196 $\pm$ 0.036 &  15.211 $\pm$ 0.039 & 15.407 $\pm$ 0.039 & 15.558 $\pm$ 0.048 &       PO \\
2017-06-11.90 & 916.40 & $ -9.61$ & 15.247 $\pm$ 0.150 & 15.076 $\pm$ 0.025 &  15.141 $\pm$ 0.050 & 15.356 $\pm$ 0.031 & 15.589 $\pm$ 0.032 &         LT \\
2017-06-12.17 & 916.67 & $ -9.35$ &                --- &                --- &  15.140 $\pm$ 0.070 & 15.270 $\pm$ 0.040 & 15.439 $\pm$ 0.065 &       PO \\
2017-06-12.20 & 916.70 & $ -9.32$ & 15.192 $\pm$ 0.072 &                --- &  15.117 $\pm$ 0.053 & 15.386 $\pm$ 0.078 & 15.525 $\pm$ 0.087 &         DN \\
2017-06-13.19 & 917.69 & $ -8.36$ &                --- & 15.034 $\pm$ 0.060 &                 --- & 15.306 $\pm$ 0.065 & 15.431 $\pm$ 0.065 &         Io \\
2017-06-13.19 & 917.69 & $ -8.36$ & 15.117 $\pm$ 0.064 &                --- &  14.895 $\pm$ 0.069 & 15.298 $\pm$ 0.079 & 15.567 $\pm$ 0.097 &         DN \\
2017-06-13.20 & 917.70 & $ -8.35$ & 15.105 $\pm$ 0.029 & 15.015 $\pm$ 0.057 &  15.029 $\pm$ 0.037 & 15.204 $\pm$ 0.069 & 15.427 $\pm$ 0.044 &       PO \\
2017-06-14.20 & 918.70 & $ -7.38$ & 15.042 $\pm$ 0.026 & 14.951 $\pm$ 0.037 &  15.012 $\pm$ 0.057 & 15.147 $\pm$ 0.044 & 15.338 $\pm$ 0.046 &       PO \\
2017-06-14.20 & 918.70 & $ -7.38$ &                --- & 14.987 $\pm$ 0.064 &                 --- & 15.195 $\pm$ 0.056 & 15.390 $\pm$ 0.067 &         Io \\
2017-06-14.20 & 918.70 & $ -7.37$ & 15.073 $\pm$ 0.072 &                --- &  14.944 $\pm$ 0.060 & 15.154 $\pm$ 0.074 & 15.406 $\pm$ 0.097 &         DN \\
2017-06-15.18 & 919.68 & $ -6.43$ &                --- & 14.902 $\pm$ 0.055 &                 --- & 15.096 $\pm$ 0.055 & 15.247 $\pm$ 0.063 &         Io \\
2017-06-15.18 & 919.68 & $ -6.43$ & 14.979 $\pm$ 0.070 &                --- &  14.903 $\pm$ 0.048 & 15.117 $\pm$ 0.068 & 15.260 $\pm$ 0.074 &         DN \\
2017-06-15.21 & 919.71 & $ -6.40$ & 14.974 $\pm$ 0.026 & 14.900 $\pm$ 0.042 &  14.945 $\pm$ 0.044 & 15.152 $\pm$ 0.067 & 15.301 $\pm$ 0.044 &       PO \\
2017-06-16.18 & 920.68 & $ -5.46$ &                --- & 14.855 $\pm$ 0.059 &                 --- & 15.073 $\pm$ 0.058 & 15.287 $\pm$ 0.074 &         Io \\
2017-06-16.18 & 920.68 & $ -5.46$ & 14.957 $\pm$ 0.078 &                --- &  14.848 $\pm$ 0.055 & 15.079 $\pm$ 0.082 & 15.312 $\pm$ 0.067 &         DN \\
2017-06-16.22 & 920.72 & $ -5.42$ & 14.925 $\pm$ 0.027 & 14.882 $\pm$ 0.028 &  14.902 $\pm$ 0.026 & 15.086 $\pm$ 0.040 & 15.262 $\pm$ 0.054 &       PO \\
2017-06-17.17 & 921.67 & $ -4.50$ &                --- & 14.796 $\pm$ 0.057 &                 --- & 14.999 $\pm$ 0.052 & 15.242 $\pm$ 0.061 &         Io \\
2017-06-17.18 & 921.68 & $ -4.49$ & 14.877 $\pm$ 0.081 &                --- &  14.776 $\pm$ 0.059 & 15.001 $\pm$ 0.087 & 15.131 $\pm$ 0.067 &         DN \\
2017-06-17.22 & 921.72 & $ -4.45$ & 14.855 $\pm$ 0.028 & 14.845 $\pm$ 0.031 &  14.884 $\pm$ 0.033 &                --- &                --- &       PO \\
2017-06-18.17 & 922.67 & $ -3.53$ & 14.830 $\pm$ 0.039 &                --- &  14.777 $\pm$ 0.030 &                --- &                --- &       PO \\
2017-06-18.18 & 922.68 & $ -3.52$ & 14.909 $\pm$ 0.085 &                --- &  14.778 $\pm$ 0.056 & 14.984 $\pm$ 0.082 & 15.188 $\pm$ 0.071 &         DN \\
2017-06-18.20 & 922.70 & $ -3.49$ &                --- & 14.798 $\pm$ 0.068 &                 --- & 14.963 $\pm$ 0.061 & 15.150 $\pm$ 0.080 &         Io \\
2017-06-20.20 & 924.70 & $ -1.56$ & 14.805 $\pm$ 0.027 & 14.720 $\pm$ 0.034 &  14.746 $\pm$ 0.039 & 14.934 $\pm$ 0.040 & 15.099 $\pm$ 0.050 &       PO \\
2017-06-21.19 & 925.69 & $ -0.60$ & 14.680 $\pm$ 0.070 &                --- &  14.770 $\pm$ 0.030 &                --- &                --- &         Ni \\
2017-06-21.22 & 925.72 & $ -0.57$ & 14.780 $\pm$ 0.028 & 14.714 $\pm$ 0.031 &  14.726 $\pm$ 0.024 & 14.911 $\pm$ 0.045 & 15.094 $\pm$ 0.042 &       PO \\
2017-06-22.22 & 926.72 & $  0.40$ & 14.799 $\pm$ 0.027 & 14.719 $\pm$ 0.032 &  14.726 $\pm$ 0.048 & 14.873 $\pm$ 0.042 & 15.041 $\pm$ 0.044 &       PO \\
2017-06-23.22 & 927.72 & $  1.37$ & 14.865 $\pm$ 0.026 & 14.765 $\pm$ 0.060 &  14.770 $\pm$ 0.040 & 14.934 $\pm$ 0.040 & 15.095 $\pm$ 0.046 &       PO \\
2017-06-24.19 & 928.69 & $  2.31$ & 14.840 $\pm$ 0.150 &                --- &  14.760 $\pm$ 0.040 &                --- &                --- &         Ni \\
2017-06-24.21 & 928.71 & $  2.34$ & 14.858 $\pm$ 0.026 & 14.763 $\pm$ 0.032 &  14.743 $\pm$ 0.025 & 14.917 $\pm$ 0.046 & 15.044 $\pm$ 0.043 &       PO \\
2017-06-25.22 & 929.72 & $  3.31$ & 14.888 $\pm$ 0.028 & 14.785 $\pm$ 0.031 &  14.761 $\pm$ 0.031 & 14.856 $\pm$ 0.019 & 15.022 $\pm$ 0.050 &       PO \\
2017-06-26.21 & 930.71 & $  4.28$ & 14.928 $\pm$ 0.030 & 14.825 $\pm$ 0.030 &  14.813 $\pm$ 0.037 & 14.960 $\pm$ 0.037 & 15.110 $\pm$ 0.046 &       PO \\
2017-06-27.22 & 931.72 & $  5.26$ & 14.929 $\pm$ 0.028 & 14.828 $\pm$ 0.030 &  14.806 $\pm$ 0.029 & 14.955 $\pm$ 0.040 & 15.098 $\pm$ 0.044 &       PO \\
2017-06-28.21 & 932.71 & $  6.22$ & 14.959 $\pm$ 0.026 & 14.853 $\pm$ 0.029 &  14.839 $\pm$ 0.038 & 14.945 $\pm$ 0.045 & 15.116 $\pm$ 0.046 &       PO \\
2017-06-29.21 & 933.71 & $  7.19$ & 14.987 $\pm$ 0.026 & 14.853 $\pm$ 0.033 &  14.893 $\pm$ 0.043 & 14.934 $\pm$ 0.040 & 15.112 $\pm$ 0.059 &       PO \\
2017-06-30.21 & 934.71 & $  8.16$ & 15.015 $\pm$ 0.030 & 14.875 $\pm$ 0.033 &  14.868 $\pm$ 0.030 & 14.955 $\pm$ 0.039 & 15.090 $\pm$ 0.043 &       PO \\
2017-07-01.21 & 935.71 & $  9.13$ & 15.011 $\pm$ 0.026 & 14.895 $\pm$ 0.026 &  14.880 $\pm$ 0.027 & 14.984 $\pm$ 0.038 & 15.120 $\pm$ 0.048 &       PO \\
2017-07-02.21 & 936.71 & $ 10.10$ & 15.030 $\pm$ 0.032 & 14.919 $\pm$ 0.030 &  14.905 $\pm$ 0.030 &                --- & 15.126 $\pm$ 0.044 &       PO \\
2017-07-03.21 & 937.71 & $ 11.07$ &                --- & 14.936 $\pm$ 0.027 &  14.913 $\pm$ 0.030 &                --- & 15.084 $\pm$ 0.044 &       PO \\
2017-07-04.21 & 938.71 & $ 12.04$ & 15.071 $\pm$ 0.030 & 14.951 $\pm$ 0.035 &  14.938 $\pm$ 0.038 & 14.992 $\pm$ 0.046 & 15.112 $\pm$ 0.047 &       PO \\
2017-07-05.21 & 939.71 & $ 13.01$ & 15.051 $\pm$ 0.032 & 14.953 $\pm$ 0.025 &  14.926 $\pm$ 0.030 & 14.961 $\pm$ 0.027 & 15.095 $\pm$ 0.051 &       PO \\
2017-07-06.21 & 940.71 & $ 13.98$ & 15.083 $\pm$ 0.031 & 14.995 $\pm$ 0.030 &  14.950 $\pm$ 0.036 & 14.994 $\pm$ 0.028 & 15.120 $\pm$ 0.045 &       PO \\
2017-07-07.21 & 941.71 & $ 14.95$ & 15.099 $\pm$ 0.036 &                --- &                 --- &                --- &                --- &       PO \\
2017-07-08.20 & 942.70 & $ 15.91$ & 15.118 $\pm$ 0.025 & 15.000 $\pm$ 0.020 &  14.974 $\pm$ 0.029 & 15.005 $\pm$ 0.078 & 15.150 $\pm$ 0.035 &       PO \\
2017-07-09.20 & 943.70 & $ 16.88$ & 15.142 $\pm$ 0.036 & 15.069 $\pm$ 0.018 &  15.025 $\pm$ 0.041 & 15.059 $\pm$ 0.036 & 15.145 $\pm$ 0.046 &       PO \\
2017-07-10.20 & 944.70 & $ 17.85$ & 15.149 $\pm$ 0.035 & 15.052 $\pm$ 0.028 &  15.022 $\pm$ 0.031 & 15.075 $\pm$ 0.039 & 15.167 $\pm$ 0.046 &       PO \\
\enddata
\end{deluxetable*}
}
{
	\begin{deluxetable*}
	{c c r c c c c l}	
\tablehead{	
	      UT Date &         JD &  Phase$ ^a $ &                  z &                  J &                  H &                  K & Telescope$ ^b $ \\
	&    2457000+  & (days) &              (mag) &              (mag) &              (mag) &              (mag) &     / Inst.    }
\startdata              
2017-05-31.90 & 905.40 & $-20.29$ & 16.641 $\pm$ 0.041 &                --- &                --- &                --- &        LT \\
2017-06-02.94 & 907.44 & $-18.30$ &                --- & 16.013 $\pm$ 0.072 & 15.555 $\pm$ 0.229 & 15.860 $\pm$ 0.250 &        NC \\
2017-06-05.94 & 910.44 & $-15.39$ & 16.255 $\pm$ 0.051 &                --- &                --- &                --- &        LT \\
2017-06-11.90 & 916.40 & $ -9.61$ & 15.903 $\pm$ 0.044 &                --- &                --- &                --- &        LT \\
2017-06-19.92 & 924.42 & $ -1.83$ & 15.604 $\pm$ 0.053 &                --- &                --- &                --- &        AF \\
2017-06-20.96 & 925.46 & $ -0.82$ &                --- & 14.852 $\pm$ 0.114 & 14.705 $\pm$ 0.261 & 14.821 $\pm$ 0.355 &        NC \\
2017-06-24.93 & 929.43 & $  3.03$ & 15.543 $\pm$ 0.067 &                --- &                --- &                --- &        AF \\
2017-07-01.89 & 936.39 & $  9.79$ & 15.402 $\pm$ 0.058 &                --- &                --- &                --- &        AF \\
2017-07-03.90 & 938.40 & $ 11.74$ &                --- & 14.700 $\pm$ 0.200 & 14.700 $\pm$ 0.250 & 14.120 $\pm$ 0.300 &        NC \\
\enddata
\end{deluxetable*}
}

\begin{deluxetable*}
	{c c r c c c c l}	
	\tablehead{	
	      UT Date &         JD &  Phase$ ^a $ &               uvw2 &               uvm2 &               uvw1 &                uvu & Telescope$ ^b $ \\
	&  2457000+  & (days) &              (mag) &              (mag) &              (mag) &              (mag) &        / Inst.   }
	\startdata              
	2017-06-02.32 & 906.82 & $-18.90$ & 14.131 $\pm$ 0.042 & 13.900 $\pm$ 0.042 & 13.991 $\pm$ 0.043 & 14.419 $\pm$ 0.045 &      UVOT \\
	2017-06-04.15 & 908.65 & $-17.12$ & 14.204 $\pm$ 0.049 & 14.015 $\pm$ 0.048 & 14.043 $\pm$ 0.052 & 14.349 $\pm$ 0.059 &      UVOT \\
	2017-06-05.18 & 909.68 & $-16.13$ & 14.131 $\pm$ 0.044 & 13.867 $\pm$ 0.042 & 13.954 $\pm$ 0.045 & 14.325 $\pm$ 0.049 &      UVOT \\
	2017-06-08.08 & 912.58 & $-13.32$ & 14.037 $\pm$ 0.050 &                --- &                --- & 14.114 $\pm$ 0.033 &      UVOT \\
	2017-06-08.41 & 912.91 & $-13.00$ & 13.999 $\pm$ 0.048 &                --- &                --- & 14.102 $\pm$ 0.033 &      UVOT \\
	2017-06-08.75 & 913.25 & $-12.67$ & 13.977 $\pm$ 0.049 &                --- &                --- & 14.039 $\pm$ 0.033 &      UVOT \\
	2017-06-10.53 & 915.03 & $-10.94$ & 13.889 $\pm$ 0.042 & 13.568 $\pm$ 0.041 & 13.639 $\pm$ 0.042 & 13.896 $\pm$ 0.042 &      UVOT \\
	2017-06-11.96 & 916.46 & $ -9.56$ & 13.845 $\pm$ 0.042 & 13.564 $\pm$ 0.042 & 13.588 $\pm$ 0.044 & 13.834 $\pm$ 0.044 &      UVOT \\
	2017-06-15.11 & 919.61 & $ -6.49$ & 13.802 $\pm$ 0.042 & 13.479 $\pm$ 0.042 & 13.419 $\pm$ 0.042 & 13.604 $\pm$ 0.042 &      UVOT \\
	2017-06-15.38 & 919.88 & $ -6.24$ & 13.778 $\pm$ 0.043 & 13.492 $\pm$ 0.042 & 13.401 $\pm$ 0.043 & 13.590 $\pm$ 0.043 &      UVOT \\
	2017-06-16.28 & 920.78 & $ -5.36$ & 13.667 $\pm$ 0.042 & 13.356 $\pm$ 0.041 & 13.303 $\pm$ 0.042 & 13.537 $\pm$ 0.043 &      UVOT \\
	2017-06-18.13 & 922.63 & $ -3.56$ & 13.677 $\pm$ 0.042 & 13.306 $\pm$ 0.042 & 13.256 $\pm$ 0.042 & 13.432 $\pm$ 0.043 &      UVOT \\
	2017-06-18.36 & 922.86 & $ -3.34$ & 13.662 $\pm$ 0.042 & 13.294 $\pm$ 0.041 & 13.207 $\pm$ 0.042 & 13.444 $\pm$ 0.041 &      UVOT \\
	2017-06-20.03 & 924.53 & $ -1.72$ & 13.732 $\pm$ 0.044 & 13.437 $\pm$ 0.042 & 13.272 $\pm$ 0.044 & 13.386 $\pm$ 0.044 &      UVOT \\
	2017-06-20.33 & 924.83 & $ -1.43$ & 13.788 $\pm$ 0.042 & 13.393 $\pm$ 0.041 & 13.274 $\pm$ 0.042 & 13.415 $\pm$ 0.041 &      UVOT \\
	2017-06-21.66 & 926.16 & $ -0.14$ & 13.930 $\pm$ 0.046 &                --- &                --- &                --- &      UVOT \\
	2017-06-22.71 & 927.21 & $  0.88$ & 13.970 $\pm$ 0.042 & 13.545 $\pm$ 0.041 & 13.379 $\pm$ 0.041 & 13.456 $\pm$ 0.039 &      UVOT \\
	2017-06-24.71 & 929.21 & $  2.82$ & 14.128 $\pm$ 0.043 & 13.706 $\pm$ 0.041 & 13.509 $\pm$ 0.042 & 13.543 $\pm$ 0.040 &      UVOT \\
	2017-06-28.53 & 933.03 & $  6.52$ & 14.438 $\pm$ 0.050 & 14.059 $\pm$ 0.046 & 13.771 $\pm$ 0.048 & 13.617 $\pm$ 0.046 &      UVOT \\
	2017-06-28.73 & 933.23 & $  6.72$ & 14.533 $\pm$ 0.052 & 14.123 $\pm$ 0.048 & 13.795 $\pm$ 0.048 & 13.594 $\pm$ 0.047 &      UVOT \\
	2017-06-30.52 & 935.02 & $  8.45$ & 14.642 $\pm$ 0.044 & 14.214 $\pm$ 0.042 & 13.884 $\pm$ 0.042 & 13.674 $\pm$ 0.038 &      UVOT \\
	2017-06-30.98 & 935.48 & $  8.91$ & 14.615 $\pm$ 0.045 & 14.278 $\pm$ 0.043 & 13.917 $\pm$ 0.043 & 13.699 $\pm$ 0.039 &      UVOT \\
	2017-07-03.38 & 937.88 & $ 11.23$ & 14.711 $\pm$ 0.042 & 14.440 $\pm$ 0.042 & 14.034 $\pm$ 0.041 & 13.755 $\pm$ 0.037 &      UVOT \\
	2017-07-04.14 & 938.64 & $ 11.97$ & 14.776 $\pm$ 0.050 & 14.492 $\pm$ 0.047 & 14.088 $\pm$ 0.047 & 13.797 $\pm$ 0.043 &      UVOT \\
	\enddata

\tablecomments{\\
	$^{a}$Rest-frame days with reference to the explosion epoch \EpEpoch.\\
$^{b}$ The abbreviations of telescope/instrument used are as follows: AS - ASAS-SN; Io - 0.5m Iowa Robotic telescope; LCO - Las Cumbres Observatory 1 m telescope
	network; LT - 2m Liverpool Telescope; DN - 0.5 m DEMONEXT telescope; PO - 0.6m telescopes of Post observatory; Ni - 1m Nickel telescope; NC - NotCAM IR imager on 2.0m NOT telescope; AF - ALFOSC mounted on 2.0m NOT telescope; UVOT - Ultraviolet optical telescope on board \textit{Swift} Satellite.\\
	Data observed within 5\,hr are represented under a single-epoch observation.}
\end{deluxetable*}

\end{center}

%% file: TR.tex
\clearpage
\startlongtable
\begin{deluxetable}
  {r c c c}
  \tablecaption{Best-fit Black-body parameters.\label{tab:BB}}

\tablehead{	 Phase$^{a}$ & Temperature $T_{\rm BB}$   & Radius $R_{\rm BB}$     &Luminosity $L_{\rm BB}$ \\
 (days)     & ($10^3$\,K) & ($10^{12}$\,m) & ($10^{44}$\,erg\,s$^{-1}$) }
\startdata
$-18.90$ & 16.90 $\pm$ 0.35 & 12.15 $\pm$ 0.52 & 0.86 $\pm$ 0.10\\
$-18.05$ & 16.45 $\pm$ 0.33 & 12.65 $\pm$ 0.51 & 0.83 $\pm$ 0.10\\
$-17.12$ & 15.83 $\pm$ 0.41 & 13.47 $\pm$ 0.73 & 0.81 $\pm$ 0.12\\
$-17.08$ & 16.12 $\pm$ 0.37 & 13.21 $\pm$ 0.64 & 0.84 $\pm$ 0.11\\
$-16.98$ & 16.04 $\pm$ 0.38 & 13.36 $\pm$ 0.68 & 0.84 $\pm$ 0.12\\
$-16.13$ & 16.08 $\pm$ 0.31 & 13.62 $\pm$ 0.56 & 0.88 $\pm$ 0.10\\
$-16.11$ & 15.87 $\pm$ 0.34 & 13.89 $\pm$ 0.64 & 0.87 $\pm$ 0.11\\
$-16.10$ & 15.96 $\pm$ 0.34 & 13.71 $\pm$ 0.59 & 0.87 $\pm$ 0.10\\
$-15.39$ & 15.76 $\pm$ 0.29 & 14.34 $\pm$ 0.54 & 0.90 $\pm$ 0.09\\
$-15.14$ & 15.77 $\pm$ 0.31 & 14.40 $\pm$ 0.58 & 0.91 $\pm$ 0.10\\
$-15.12$ & 15.83 $\pm$ 0.28 & 14.29 $\pm$ 0.53 & 0.91 $\pm$ 0.09\\
$-14.18$ & 15.61 $\pm$ 0.33 & 15.13 $\pm$ 0.67 & 0.97 $\pm$ 0.12\\
$-14.15$ & 15.67 $\pm$ 0.29 & 15.00 $\pm$ 0.57 & 0.97 $\pm$ 0.10\\
$-13.32$ & 15.44 $\pm$ 0.30 & 15.83 $\pm$ 0.63 & 1.02 $\pm$ 0.11\\
$-13.00$ & 15.55 $\pm$ 0.35 & 15.80 $\pm$ 0.73 & 1.04 $\pm$ 0.13\\
$-12.67$ & 15.43 $\pm$ 0.30 & 16.24 $\pm$ 0.69 & 1.06 $\pm$ 0.12\\
$-12.47$ & 15.43 $\pm$ 0.31 & 16.31 $\pm$ 0.72 & 1.07 $\pm$ 0.13\\
$-12.30$ & 15.32 $\pm$ 0.28 & 16.64 $\pm$ 0.65 & 1.09 $\pm$ 0.12\\
$-12.23$ & 15.36 $\pm$ 0.31 & 16.57 $\pm$ 0.70 & 1.09 $\pm$ 0.13\\
$-11.29$ & 15.21 $\pm$ 0.26 & 17.34 $\pm$ 0.66 & 1.15 $\pm$ 0.12\\
$-11.28$ & 15.23 $\pm$ 0.29 & 17.28 $\pm$ 0.72 & 1.15 $\pm$ 0.13\\
$-11.26$ & 15.17 $\pm$ 0.25 & 17.45 $\pm$ 0.60 & 1.15 $\pm$ 0.11\\
$-10.94$ & 15.13 $\pm$ 0.27 & 17.76 $\pm$ 0.66 & 1.18 $\pm$ 0.12\\
$-10.32$ & 14.98 $\pm$ 0.27 & 18.27 $\pm$ 0.69 & 1.20 $\pm$ 0.12\\
$-10.31$ & 14.92 $\pm$ 0.24 & 18.44 $\pm$ 0.66 & 1.20 $\pm$ 0.12\\
$-10.29$ & 14.99 $\pm$ 0.23 & 18.25 $\pm$ 0.62 & 1.20 $\pm$ 0.11\\
$ -9.58$ & 14.86 $\pm$ 0.29 & 18.79 $\pm$ 0.80 & 1.23 $\pm$ 0.14\\
$ -9.35$ & 14.69 $\pm$ 0.28 & 19.34 $\pm$ 0.81 & 1.24 $\pm$ 0.14\\
$ -9.32$ & 14.70 $\pm$ 0.26 & 19.31 $\pm$ 0.76 & 1.24 $\pm$ 0.13\\
$ -8.36$ & 14.34 $\pm$ 0.22 & 20.73 $\pm$ 0.69 & 1.29 $\pm$ 0.12\\
$ -8.35$ & 14.35 $\pm$ 0.23 & 20.70 $\pm$ 0.71 & 1.29 $\pm$ 0.12\\
$ -7.38$ & 14.16 $\pm$ 0.19 & 21.70 $\pm$ 0.69 & 1.35 $\pm$ 0.11\\
$ -7.37$ & 14.15 $\pm$ 0.20 & 21.72 $\pm$ 0.67 & 1.35 $\pm$ 0.11\\
$ -6.49$ & 13.90 $\pm$ 0.19 & 22.95 $\pm$ 0.72 & 1.40 $\pm$ 0.12\\
$ -6.43$ & 14.00 $\pm$ 0.19 & 22.80 $\pm$ 0.66 & 1.42 $\pm$ 0.11\\
$ -6.40$ & 14.03 $\pm$ 0.20 & 22.70 $\pm$ 0.72 & 1.42 $\pm$ 0.12\\
$ -6.24$ & 13.89 $\pm$ 0.21 & 23.09 $\pm$ 0.81 & 1.42 $\pm$ 0.13\\
$ -5.46$ & 14.06 $\pm$ 0.23 & 23.25 $\pm$ 0.90 & 1.50 $\pm$ 0.15\\
$ -5.42$ & 14.07 $\pm$ 0.20 & 23.22 $\pm$ 0.72 & 1.51 $\pm$ 0.13\\
$ -5.36$ & 14.24 $\pm$ 0.20 & 22.91 $\pm$ 0.71 & 1.54 $\pm$ 0.13\\
$ -4.50$ & 13.98 $\pm$ 0.20 & 24.09 $\pm$ 0.80 & 1.58 $\pm$ 0.14\\
$ -4.49$ & 13.96 $\pm$ 0.18 & 24.17 $\pm$ 0.71 & 1.58 $\pm$ 0.12\\
$ -4.45$ & 14.01 $\pm$ 0.19 & 24.00 $\pm$ 0.69 & 1.58 $\pm$ 0.12\\
$ -3.56$ & 13.84 $\pm$ 0.17 & 24.87 $\pm$ 0.72 & 1.62 $\pm$ 0.12\\
$ -3.53$ & 13.82 $\pm$ 0.19 & 24.99 $\pm$ 0.77 & 1.62 $\pm$ 0.13\\
$ -3.52$ & 13.90 $\pm$ 0.24 & 24.61 $\pm$ 1.01 & 1.61 $\pm$ 0.17\\
$ -3.49$ & 13.84 $\pm$ 0.21 & 24.87 $\pm$ 0.89 & 1.62 $\pm$ 0.15\\
$ -3.34$ & 13.89 $\pm$ 0.20 & 24.89 $\pm$ 0.84 & 1.64 $\pm$ 0.14\\
$ -1.78$ & 13.24 $\pm$ 0.16 & 26.93 $\pm$ 0.74 & 1.59 $\pm$ 0.12\\
$ -1.50$ & 13.19 $\pm$ 0.15 & 27.10 $\pm$ 0.68 & 1.59 $\pm$ 0.11\\
$ -0.82$ & 12.95 $\pm$ 0.14 & 27.89 $\pm$ 0.68 & 1.56 $\pm$ 0.10\\
$ -0.57$ & 12.86 $\pm$ 0.15 & 28.22 $\pm$ 0.74 & 1.55 $\pm$ 0.11\\
$ -0.14$ & 12.69 $\pm$ 0.15 & 28.69 $\pm$ 0.75 & 1.52 $\pm$ 0.11\\
$  0.40$ & 12.62 $\pm$ 0.14 & 28.80 $\pm$ 0.73 & 1.50 $\pm$ 0.10\\
$  0.88$ & 12.58 $\pm$ 0.13 & 28.67 $\pm$ 0.72 & 1.47 $\pm$ 0.10\\
$  1.37$ & 12.52 $\pm$ 0.14 & 28.51 $\pm$ 0.77 & 1.42 $\pm$ 0.10\\
$  2.34$ & 12.26 $\pm$ 0.13 & 29.22 $\pm$ 0.74 & 1.37 $\pm$ 0.09\\
$  2.82$ & 12.19 $\pm$ 0.11 & 29.15 $\pm$ 0.59 & 1.34 $\pm$ 0.07\\
$  3.03$ & 12.11 $\pm$ 0.13 & 29.43 $\pm$ 0.75 & 1.33 $\pm$ 0.09\\
$  3.31$ & 12.01 $\pm$ 0.11 & 29.76 $\pm$ 0.66 & 1.31 $\pm$ 0.08\\
$  4.28$ & 11.83 $\pm$ 0.10 & 29.93 $\pm$ 0.64 & 1.25 $\pm$ 0.07\\
$  5.26$ & 11.52 $\pm$ 0.12 & 31.01 $\pm$ 0.75 & 1.21 $\pm$ 0.08\\
$  6.22$ & 11.30 $\pm$ 0.10 & 31.62 $\pm$ 0.67 & 1.16 $\pm$ 0.06\\
$  6.52$ & 11.32 $\pm$ 0.11 & 31.33 $\pm$ 0.75 & 1.15 $\pm$ 0.07\\
$  6.72$ & 11.08 $\pm$ 0.10 & 32.59 $\pm$ 0.73 & 1.14 $\pm$ 0.07\\
$  7.19$ & 11.10 $\pm$ 0.10 & 32.14 $\pm$ 0.75 & 1.12 $\pm$ 0.07\\
$  8.16$ & 10.95 $\pm$ 0.10 & 32.50 $\pm$ 0.71 & 1.08 $\pm$ 0.06\\
$  8.45$ & 10.90 $\pm$ 0.08 & 32.66 $\pm$ 0.56 & 1.07 $\pm$ 0.05\\
$  8.91$ & 10.88 $\pm$ 0.08 & 32.49 $\pm$ 0.61 & 1.05 $\pm$ 0.05\\
$  9.13$ & 10.83 $\pm$ 0.09 & 32.76 $\pm$ 0.62 & 1.05 $\pm$ 0.05\\
$  9.79$ & 10.76 $\pm$ 0.07 & 32.82 $\pm$ 0.59 & 1.03 $\pm$ 0.05\\
\enddata
\tablecomments{\\
	Temperature and radius are estimated from black-body fitting, while luminosities are computed from fitted parameters.\\
	$^{a}$Rest-frame days relative to the epoch of the $g$-band peak at \EpEpoch.\\
	}
\end{deluxetable}

%% file: speclog.tex

\begin{table*}
\centering
  \caption{Summary of optical spectroscopy of \sn.}
  \label{tab:speclog}
  \begin{tabular}{lc cl cc cc}
    \hline
 UT Date          & JD $-$         & Phase$^{a}$& Instrument & Exposure  &  Slit width   &  Wavelength range  & Resolution                \\
               & 2,457,900   & (days)     &            & (s)    &  ($''$)    &  (\AA)             & ($\lambda/\Delta\lambda$) \\ \hline
 2017-05-30.9  & 04.41      & -21.2      & NOT/ALFOSC & 1800     &  1.0          &  3200-9450         & 330                       \\
 2017-05-31.9$^{b}$  & 05.38      & -20.3      & LT/SPRAT   & 350      &  1.8          &  4000-8000         & 350                       \\
 2017-06-03.2  & 07.70      & -18.1      & Shane/Kast & 900      &  2.0          &  3300-10600        & 680                        \\
 2017-06-05.9$^{b}$  & 10.41      & -15.4      & LT/SPRAT   & 350      &  1.8          &  4000-8000         & 350                       \\
 2017-06-08.0  & 12.48      & -13.4      & NOT/ALFOSC & 1800     &  1.0          &  3200-9450         & 330                       \\
 2017-06-10.9$^{b}$  & 15.41      & -10.6      & Asiago/B\&C  & 1200     &  2.2          &  3400-9200         & 700                       \\
 2017-06-10.9  & 15.42      & -10.6      & LT/SPRAT   & 400      &  1.8          &  4000-8000         & 350                       \\
 2017-06-11.9  & 16.38      & -9.6       & Asiago/B\&C  & 1800     &  2.2          &  3400-9200         & 700                       \\
 2017-06-15.9  & 20.39      & -5.7       & LT/SPRAT   & 400      &  1.8          &  4000-8000         & 350                       \\
 2017-06-17.2  & 21.68      & -4.5       & FLWO/FAST  & 900      &  1.5          &  3500-8000         & 1800                       \\
 2017-06-17.9  & 22.43      & -3.9       & NOT/ALFOSC & 1800     &  1.0          &  3200-9450         & 330                       \\
 2017-06-19.0  & 23.50      & -2.7       & Shane/AeroSpOpIR & 3600       &  1.1           &  5000-22000         & 600                        \\
 2017-06-20.2  & 24.70      & -1.6       & Shane/Kast & 600      &  2.0          &  3300-10600        & 690                       \\
 2017-06-20.9  & 25.43      & -0.9       & LT/SPRAT   & 400      &  1.8          &  4000-8000         & 350                       \\
 2017-06-21.0$^{b}$  & 25.46      & -0.8       & NOT/NotCam & 3600     &  1.6           &  10000-13000        & 500                        \\
 2017-06-21.2$^{c}$  & 25.67      & -0.6       & FLWO/FAST  & 900      &  3.0          &  3500-8000         & 1800                      \\
 2017-06-21.2  & 25.70      & -0.6       & Shane/Kast & 1800     &  2.0          &  3300-10600        & 690                        \\
 2017-06-21.3  & 25.79      & -0.5       & IRTF/SpeX &   540    &  0.5          &  8000-24000        & 100                        \\
 2017-06-23.2  & 27.70      &  1.4       & Shane/Kast & 1200     &  1.5          &  3600-8200         & 1300                      \\
 2017-06-24.7  & 28.71      &  2.3       & Shane/Kast & 1500     &  2.0          &  3500-8800         & 950                       \\
 2017-06-24.9  & 29.41      &  3.0       & NOT/ALFOSC & 1100     &  1.0          &  3200-9450         & 330                       \\
 2017-06-25.2  & 29.68      &  3.3       & Shane/Kast & 1800     &  2.0          &  3300-9000         & 710                       \\
 2017-06-26.2  & 30.70      &  4.3       & Shane/Kast & 1200     &  2.0          &  3300-10600        & 680                       \\
 2017-06-26.9  & 31.43      &  5.0       & NOT/ALFOSC & 1800     &  1.0          &  3200-9450         & 330                       \\
 2017-06-27.2  & 31.69      &  5.2       & Shane/Kast & 1442     &  2.0          &  3300-10600        & 690                       \\
 2017-06-28.2  & 32.71      &  6.2       & Shane/Kast & 1800     &  2.0          &  3600-10600        & 710                       \\
 2017-07-01.2  & 35.69      &  9.1       & Shane/Kast & 750      &  2.0          &  3300-10600        & 690                       \\
 2017-07-01.9  & 36.41      &  9.8       & NOT/ALFOSC & 1200     &  1.3          &  3200-9450         & 250                       \\
 2017-07-05.2  & 39.69      & 13.0       & Shane/Kast & 900      &  2.0          &  3400-7800         & 960                       \\
 2017-07-12.9$^{b}$  & 47.36      & 21.1       & Asiago/B\&C  & 1200     &  2.2          &  3400-9200         & 700                       \\
 2017-07-18.2  & 52.68      & 25.6       & Shane/Kast & 360      &  2.0          &  3300-10600        & 690                       \\
 2017-07-18.9  & 53.35      & 26.2       & Asiago/AFOSC& 1200    &  1.69         &  3360-7740         & 360                        \\

    \hline
  \end{tabular}
\begin{flushleft}
  $^{a}$Relative to the \textit{g}-band maximum on \EpEpoch.\\
  $^{b}$These spectrum is not shown in the figures due to their low SNR.\\
  $^{c}$This spectrum is used to estimate the host-galaxy properties, but not shown in the figures.\\
\end{flushleft}
\end{table*}